\documentclass[a4paper,twocolumn,11pt,accepted=2023-01-05]{quantumarticle}
\pdfoutput=1
\usepackage[utf8]{inputenc}
\usepackage[english]{babel}
\usepackage[T1]{fontenc}
\usepackage{amsmath}
\usepackage{hyperref}
\usepackage{textcomp}
\usepackage{tikz}
\usepackage{lipsum}
\usepackage{graphicx}
\usepackage[section]{placeins}
\usepackage[numbers,sort&compress]{natbib}

\begin{document}

	\title{Generation of highly-retrievable atom–photon entanglement with a millisecond lifetime via a spatially-multiplexed cavity}
	
	\author{Minjie Wang}
	\email{wangminjie2024@163.com}
	\affiliation{The State Key Laboratory of Quantum Optics and Quantum Optics Devices, Institute of Opto-Electronics, Shanxi University, Taiyuan 030006 }
	\affiliation{China Collaborative Innovation Center of Extreme Optics,
		Shanxi University, Taiyuan 030006, China }
	\author{Shengzhi Wang}
    \email{wangshengzhi5236@163.com}
	\affiliation{The State Key Laboratory of Quantum Optics and Quantum Optics Devices, Institute of Opto-Electronics, Shanxi University, Taiyuan 030006 }
	\affiliation{China Collaborative Innovation Center of Extreme Optics,
		Shanxi University, Taiyuan 030006, China }
	\author{Tengfei Ma}
	\email{527041358@qq.com}
	\affiliation{The State Key Laboratory of Quantum Optics and Quantum Optics Devices, Institute of Opto-Electronics, Shanxi University, Taiyuan 030006 }
	\affiliation{China Collaborative Innovation Center of Extreme Optics,
		Shanxi University, Taiyuan 030006, China }

	\author{Ya Li}
	\email{liya3642@163.com}
	\affiliation{The State Key Laboratory of Quantum Optics and Quantum Optics Devices, Institute of Opto-Electronics, Shanxi University, Taiyuan 030006 }
	\affiliation{China Collaborative Innovation Center of Extreme Optics,
		Shanxi University, Taiyuan 030006, China }
	
	\author{Yan Xie}
	\email{x2454900109@163.com}
	\affiliation{The State Key Laboratory of Quantum Optics and Quantum Optics Devices, Institute of Opto-Electronics, Shanxi University, Taiyuan 030006 }
	\affiliation{China Collaborative Innovation Center of Extreme Optics,
		Shanxi University, Taiyuan 030006, China }
	
	\author{Haole Jiao}
	\email{15340989229@163.com}
	\affiliation{The State Key Laboratory of Quantum Optics and Quantum Optics Devices, Institute of Opto-Electronics, Shanxi University, Taiyuan 030006 }
	\affiliation{China Collaborative Innovation Center of Extreme Optics,
		Shanxi University, Taiyuan 030006, China }

	\author{Hailong Liu}
	\email{18903523250@189.cn}
	\affiliation{The State Key Laboratory of Quantum Optics and Quantum Optics Devices, Institute of Opto-Electronics, Shanxi University, Taiyuan 030006 }
	\affiliation{China Collaborative Innovation Center of Extreme Optics,
		Shanxi University, Taiyuan 030006, China }
	
	\author{Shujing Li}
	\email{lishujing@sxu.edu.cn}
	\affiliation{The State Key Laboratory of Quantum Optics and Quantum Optics Devices, Institute of Opto-Electronics, Shanxi University, Taiyuan 030006 }
	\affiliation{China Collaborative Innovation Center of Extreme Optics,
		Shanxi University, Taiyuan 030006, China }

	\author{Hai Wang}
	\email{wanghai@sxu.edu.cn}
	\affiliation{The State Key Laboratory of Quantum Optics and Quantum Optics Devices, Institute of Opto-Electronics, Shanxi University, Taiyuan 030006 }
	\affiliation{China Collaborative Innovation Center of Extreme Optics,
		Shanxi University, Taiyuan 030006, China }

	\maketitle
	
	\begin{abstract}
		Qubit memory entangled with photonic qubit is the building block for quantum repeaters.Cavity-enhanced spin-wave–photon entanglement with sub-second lifetime has been demonstrated by applying dual control modes onto optical-lattice atoms.However, owing to double-mode retrievals in that experiment, retrieval efficiency of spin-wave qubit is about one quarter lower than that for a single spin-wave mode.Here, by coupling cold atoms to two modes of a polarization-interferometer-based cavity,we achieve perfectly-enhanced qubit retrieval in long-lived atom-photon entanglement.A write-laser beam is applied onto the cold atoms to create a magnetic-field-insensitive spin-wave qubit entangled with a photonic qubit.The spin-wave qubit is retrieved with a single-mode read-laser beam, and the quarter retrieval-efficiency loss is avoided.Our experimental data shows that zero-delay intrinsic retrieval efficiency is up to 77\% and 1/e lifetime 1ms. At 50\% retrieval efficiency, the storage time reaches 540$\mu$s, which is 13.5 times longer than the best reported result.
	\end{abstract}
        \section{Introduction}
       A qubit memory entangled with a photonic qubit forms the building block (repeater node) for long-distance quantum communications \cite{1,2,3}, large-scale quantum internet \cite{4, 5}, and quantum clock network \cite{6}  through a quantum repeater (QR) \cite{1, 2}. In past decades, optical quantum memories (QMs) have been experimentally demonstrated with various matter systems \cite{7} such as atomic ensembles \cite{1, 8} and single-quantum systems including individual atoms \cite{9, 10}, ions \cite{11}, and solid-state spins \cite{12, 13}. Compared with the QMs in a single-quantum system, the atomic-ensemble-based QMs feature collective enhancement and promise higher retrieval efficiencies \cite{1}.With atomic ensembles, optical QMs has been demonstrated via various schemes \cite{7, 8, 14, 15}including Duan–Lukin–Cirac–Zoller (DLCZ) protocol \cite{1,16,17,18,19,20,21,22,23,24,25,26,27,28,29,30,31,32,33,34,35,36,37,38} and “read-write” \cite{14} (or called “absorptive” \cite{15}) schemes such as electromagnetically-induced-transparency (EIT) dynamics \cite{39,40,41,42,43,44,45,46,47,48,49,50,51,52}, photon echoes \cite{53,54,55,56,57,58,59,60}, Raman memory \cite{61,62,63} and its variants \cite{64, 65}, etc. In contrast to the “read-write” schemes, the DLCZ protocol establishes QMs via spontaneous Raman emissions of Stokes photon induced by a write pulse instead of storing single specific (incoming) photons \cite{14}. Since DLCZ protocol can directly produces non-classically correlated \cite{16,17,18,19,20,21,22,23,24,25,26,27,28} or entangled \cite{29,30,31,32,33,34,35,36,37,38} pairs of a Stokes photon and a spin-wave memory, it benefits for practically realizing QRs \cite{1}.
       
       High-efficiency and long-lived QMs are required for effectively achieving quantum information tasks \cite{1, 3, 7}. For the distribution of an entangled pair over a distance L =600 km using quantum repeater based on atomic-ensemble memories that have enough long lifetime, Ref. \cite{1} plots the required average time as a function of the memory efficiency in Fig. 19. 
       The Figure shows that the efficiency of the memories is enormous importance to achieve reasonable entanglement distribution rates. For example, an increase in the memory efficiency from 89\% to 90\%, i.e., a ~1\% increase in retrieval efficiency, leads to a decrease (an improvement) in the entanglement distribution time (rate) by 10\%–14\% (11\%-16\%), depending on the protocol. For a practical memory, the lifetime at 50\% efficiency denotes the storage time within which one can retrieve the memory with an efficiency exceeding 50\%.
       The lifetime at 50\% efficiency has been used to characterize high-performance memories that have simultaneously high retrieval efficiency and long lifetime in previous works \cite{20, 24}.
       Additionally, establishing entanglement between two nodes separated by L=300 km in a ‘‘heralded’’ fashion requires storing a qubit for at least L/c=1.5 ms \cite{34, 66}, with c being the speed of light in fibers. To realize long-lived QMs, significant progresses have been made with cold atomic ensembles \cite{19,20,21,22,23,24, 32,33,34, 56}. These studies show that atomic-motion-induced decoherence can be suppressed either by lengthening spin-wave wavelengths \cite{20, 21, 32, 34, 55} or confining the atoms in optical lattices \cite{22,23,24, 32,33}.
       Inhomogeneous-broadening-induced decoherence may be suppressed by storing spin waves (SWs) in magnetic-field-insensitive (MFI) coherences, which includes three spin transitions $\left| {5{S_{{\rm{1/2}}}},F = {\rm{1, }}{m_F} =  \pm {\rm{1,0}}} \right\rangle  \leftrightarrow \left| {5{S_{{\rm{1/2}}}},F = {\rm{2,}}{m_F} =  \mp {\rm{1}},{\rm{0}}} \right\rangle $ for $^{87}$Rb atoms, where,${m_{{F_1}}} = {\rm{0}} \leftrightarrow {m_{{F_2}}}{\rm{ = 0}}$ is called clock coherence. In a specific DLCZ experiment, one can store SWs in one of the MFI coherences \cite{24}.
       
       In previous long-lived atom–photon entanglement generation via DLCZ protocol \cite{32}, a Mach–Zehnder interferometer, whose two arms are used to encode photonic qubit, is built around optical-lattice Rb atoms. Two spatially-distinctive SWs associated with the clock coherence and correlated with the Stokes fields emitting into the two arms, respectively, are created by a write beam and stored as memory qubit.
       To maintain maximal entanglement, the relative phase between the two arms was actively stabilized by coupling an auxiliary laser beam into the interferometer.The lifetime of entanglement storage reaches 0.1 s but the retrieval efficiency is only 16\% due to weak atom-photon interactions \cite{32}. The high-efficiency memories for single photons have been achieved by using either high-optical-depth cold atoms \cite{50, 51} or coupling moderate-optical-depth atoms with a low-finesse optical cavity \cite{21, 24, 30, 33}.In high-optical-depth cold atoms, the efficiencies of the storages of single-photon entanglement \cite{51} and polarization qubit \cite{50} reach 85\%, while their lifetimes are only $\sim$15$\mu$s.
       Using the cavity-enhanced scheme, Pan’s group demonstrated intrinsic retrieval efficiency up to 76\% in a DLCZ experiment that create polarization atom-photon entanglement \cite{30}.In that experiment, the spin-wave qubit is stored as superimposition of magnetic-field-sensitive and MFI coherences. Limiting to fast decoherence of the magnetic-field-sensitive coherence, the memory lifetime is only 30 µs \cite{30}. Efficient and long-lived atom-photon quantum correlations has been demonstrated by applying a write beam onto optical-lattice atoms, where the atomic SW is stored in the clock coherence and coupled to a ring cavity with its length being locked \cite{24}. The efficiencies at zero delay reach to 76\% and at 50-ms storage time to 50\%.
       \cite{24}.On this basis, Pan’s group demonstrated cavity-enhanced atom-photon entanglement with sub-second lifetime\cite{33}.The group applied two spatially-distinctive write beams, one of which is H-polarized and another V-polarized, onto the lattice atoms to create H- and V- polarized Stokes emissions into single-mode cavity. The two polarizations of the Stokes emissions form photonic qubit.At the same time, two “clock” SW modes, one is correlated with H-polarized and another with V-polarized Stokes fields, are created and stored as memory qubit.
       That experiment \cite{33} removes the M–Z interferometer used in Ref. \cite{32} and then avoids experimental complexity due to interferometer-mediated coupling between atoms and a cavity. The SW qubit is retrieved by applying H- and V- polarized read beams, where the H (V) read beam is applied along the direction opposite to the H- (V-) polarized write beam. The SW correlated with the H (V) Stokes photon is mapped into the cavity mode by the H- (V-) polarized read beam. However, the SW correlated with the H (V) photon is also retrieved by the V- (H-) polarized read-laser beam and direct into the directions other than the cavity mode. Such double-mode retrievals for SW qubit \cite{33} leads to a significant decrease in retrieval efficiency compared to the single-mode retrieval in Ref. \cite{24} at all storage times.
       For example, the efficiency of retrieving the spin-wave qubit was only 58\% at zero delay \cite{33}, about three quarters that of retrieving the single-mode one (76\%) \cite{24}.At 50\% efficiency, the lifetime of the memory for qubit was only $\sim$40$\mu$s \cite{33}.Here, we overcome the imperfect retrieval in Ref.\cite{33} by coupling cold atoms to a two-mode ring cavity based on a polarization interferometer. The interferometer is mainly formed by two beam displacers (BD1 and BD2). The relative phase between the two arms is passively stabilized, which has been demonstrated in previous experiments \cite{43, 49,50,51, 67,68}.Two optical lenses are inserted in the interferometer, which make interferometer’s two arms (A$_R$ and A$_L$) crossway pass-through cold atoms.
       By arranging interferometer’s configuration, the ring cavity simultaneously supports the two arms as cavity modes cf. below. By applying a write beam, we create two SWs associated with the clock coherence and correlated with the Stokes emissions into the cavity modes A$_R$ and A$_L$, respectively.The two SWs are retrieved by a read beam and then avoid the double-mode retrieval.The two Stokes (retrieved) fields form a photonic qubit and resonate with the cavity. The intrinsic qubit retrieval efficiencies reach 50\% (66.7\%) at 540-µs (230-µs) storage time, which is 13.5 (24) times higher than the best reported results \cite{33} (\cite{30}).The measured Bell parameter for atom-photon entanglement is ${\rm{2}}{\rm{.5}} \pm {\rm{0}}{\rm{.02}}$  at zero delay.Unlike the previous experiments \cite{21, 24, 30, 33, 69,70}, our experiment demonstrated that the two arms of the polarization interferometer can be simultaneously supported by a ring cavity, which enable perfect retrieval in cavity-enhanced and long-lived atom–photon entanglement.
       \section{Experimental method}
       As shown in the schematic diagram Fig.1a, the heart of experimental setup is a ring cavity inserted by a polarization interferometer with cold atoms (circled by dashed line). The ring cavity is formed by three flat mirrors (HR$_{1,2,3}$) with high reflection and a flat output coupler (OC) with a reflectance of 80\%. It supports TEM$_{00}$ mode A$_{00}$ that propagates in the whole cavity [71]. When the polarization interferometer is inserted in the cavity, the  A$_{00}$  mode has the path from BD2 to BD1 via OC and HR$_1$ (black line in Fig.1a). When the mode  A$_{00}$  propagates along clockwise, it will be split into H (horizontally) and V (vertically) -polarized components by BD2, both direct into the interferometer’s two arms A$_R$ and A$_L$, respectively. Two optical lenses lens1 and lens2, which have the same focus length F$_0$, are inserted in the interferometer. Then, the arms A$_R$ and A$_L$ crossway pass through the atoms and couple with the atoms, respectively. The more detailed explanations of the A$_R$ and A$_L$ paths can be found in Supplementary material \cite{71}. By setting the distance between lens1 and lens2 to be 2F$_0$, the two arms are well recombined into the mode A$_{00}$ by BD1 and then supported by the cavity. In our experiment, the spot size of A$_{00}$ is set to be a large value (5.2mm) in order to decrease the mode losses escaping from the cavity. The measured loss of the mode A$_{00}$ via A$_R$ (A$_L$)  path escaping from cavity per round trip is $\sim$3.2\% (3.2\%). So, the two arms A$_R$ and A$_L$ may serve as cavity modes.
       
       The atomic ground levels$\left| a \right\rangle  = \left| {5{S_{1/2}},F{\rm{ = 1}}} \right\rangle $ and $\left| b \right\rangle  = \left| {5{S_{1/2}},F{\rm{ = 2}}} \right\rangle $  together with the excited level$\left| {{e_{\rm{1}}}} \right\rangle  = \left| {5{P_{1/2}},F'{\rm{ = 1}}} \right\rangle $  ($\left| {{e_{\rm{2}}}} \right\rangle  = \left| {5{P_{1/2}},F'{\rm{ = 2}}} \right\rangle $) form a $\Lambda $-type system [Fig.1b and c].
       After the atoms are prepared in the Zeeman state $\left| {a,{m_{{F_a}}} = 0} \right\rangle $, we start spin-wave-photon (atom-photon) entanglement generation. At the beginning of a trial \cite{71}, a 795-nm ${\sigma ^{\rm{ + }}}$ -polarized write pulse with red-detuned by 110 MHz to the $\left| a \right\rangle  \to \left| {{e_{\rm{1}}}} \right\rangle $ transition is applied to the atoms along z-axis through a beam splitter BS1. This write pulse induces the Raman transition $\left| {a,{m_{Fa}} = {\rm{0}}} \right\rangle  \to \left| {b,{m_{Fb}} = {\rm{0}}} \right\rangle $ via $\left| {{e_1},{m_{Fe}} = {\rm{1}}} \right\rangle $ [Fig. 1b], which emit  ${\sigma ^{\rm{ + }}}$ -polarized Stokes photons and simultaneously create SW excitations associated with the clock coherence $\left| {{m_a} = {\rm{0}}} \right\rangle  \leftrightarrow \left| {{m_b} = {\rm{0}}} \right\rangle $.
       If the Stokes photon is emitted into the cavity mode A$_R$ (A$_L$) and moves towards the right, one collective excitation will be created in the SW mode M$_R$ (M$_L$) defined by the wave-vector $k_{{M_R}}^{} = {k_W} - k_{{S_R}}^{}$ ($k_{{M_L}}^{} = {k_W} - k_{{S_L}}^{}$), where ${k_W}$ is the wave-vector of the write pulse and $k_{{S_R}}^{}$ ($k_{{S_L}}^{}$) that of the Stokes photon in the cavity  A$_R$ (A$_L$) mode. The angle between the mode  A$_R$ (A$_L$) and the write beam is ${\theta _R} \approx {\rm{0}}{\rm{.05}}{{\rm{3}}^{\rm{0}}}$ ($[{\theta _l} \approx  - {\rm{0}}{\rm{.05}}{{\rm{3}}^{\rm{0}}}$) \cite{71}. Such small angles make SW wavelengths be long, suppressing atomic-motion-induced decoherence \cite{34}.The ${\sigma ^{\rm{ + }}}$ -polarized Stokes fields in A$_R$ and A$_L$ modes propagate along clockwise, which are transformed into H-polarized fields by a $\lambda$/4 wave-plate QW$_S$. Furthermore, the H-polarized field in  A$_L$ is transformed into V-polarized field by a $\lambda$/2 wave-plate HW$_S$.
       Both fields are combined into the cavity mode A$_{00}$ by BD2 and form a Stokes qubit ${S_{qbit}}$. As shown in Fig. 1(b), the write pulse also induces the Raman transition $\left| {a,{m_{Fa}} = {\rm{0}}} \right\rangle  \to \left| {b,{m_{Fb}} = {\rm{2}}} \right\rangle $ via $\left| {{e_1},{m_{Fe}} = {\rm{1}}} \right\rangle $ , which emit ${\sigma ^ - }$ -polarized Stokes photons and simultaneously create SWs associated with the magnetic-field-sensitive coherence $\left| {{m_{Fa}} = {\rm{0}}} \right\rangle  \leftrightarrow \left| {{m_{Fb}} = {\rm{2}}} \right\rangle $. If the ${\sigma ^ - }$-polarized Stokes photon direct into the mode A$_R$ (A$_L$) and propagate along clockwise, it will be transformed into V- (H-) polarized photon by the QW$_S$ (QW$_S$ and HW$_S$) and then is excluded from the A$_{00}$ cavity mode by BD2. Propagating in A$_{00}$, the  ${S_{qbit}}$ returns the interferometer. It is split into H and V -polarized fields by BD1, which direct into  A$_R$ and A$_L$  modes, respectively. Both Stokes modes are transformed into ${\sigma ^{\rm{ + }}}$ -polarization by wave-plates \cite{71} and then interact with the atoms again. The spin-wave qubit formed by M$_R$ and M$_L$ modes is entangled with ${S_{qbit}}$ , which is written as:
       \begin{equation}
       	\Phi _{{\rm{a - p}}}^{} = \left( {\left| H \right\rangle _S^{}\left| {{M_R}} \right\rangle  + {e^{i{\varphi _{_S}}}}\left| V \right\rangle _S^{}\left| {{M_L}} \right\rangle } \right)/\sqrt 2
       \end{equation}
       where, $\left| H \right\rangle _S^{}$ ($\left| V \right\rangle _S^{}$) denotes the H- (V-) polarized Stokes photon, $\left| {{M_R}} \right\rangle $ ($\left| {{M_L}} \right\rangle $) one SW excitation in the mode M$_R$ ( M$_L$), ${\varphi _S}$ the relative phase between the two Stokes emissions from the atoms to BD2 via  A$_R$ and A$_L$ paths.
       
     \begin{figure}[h]
     	\centering
     	\includegraphics[width=3in]{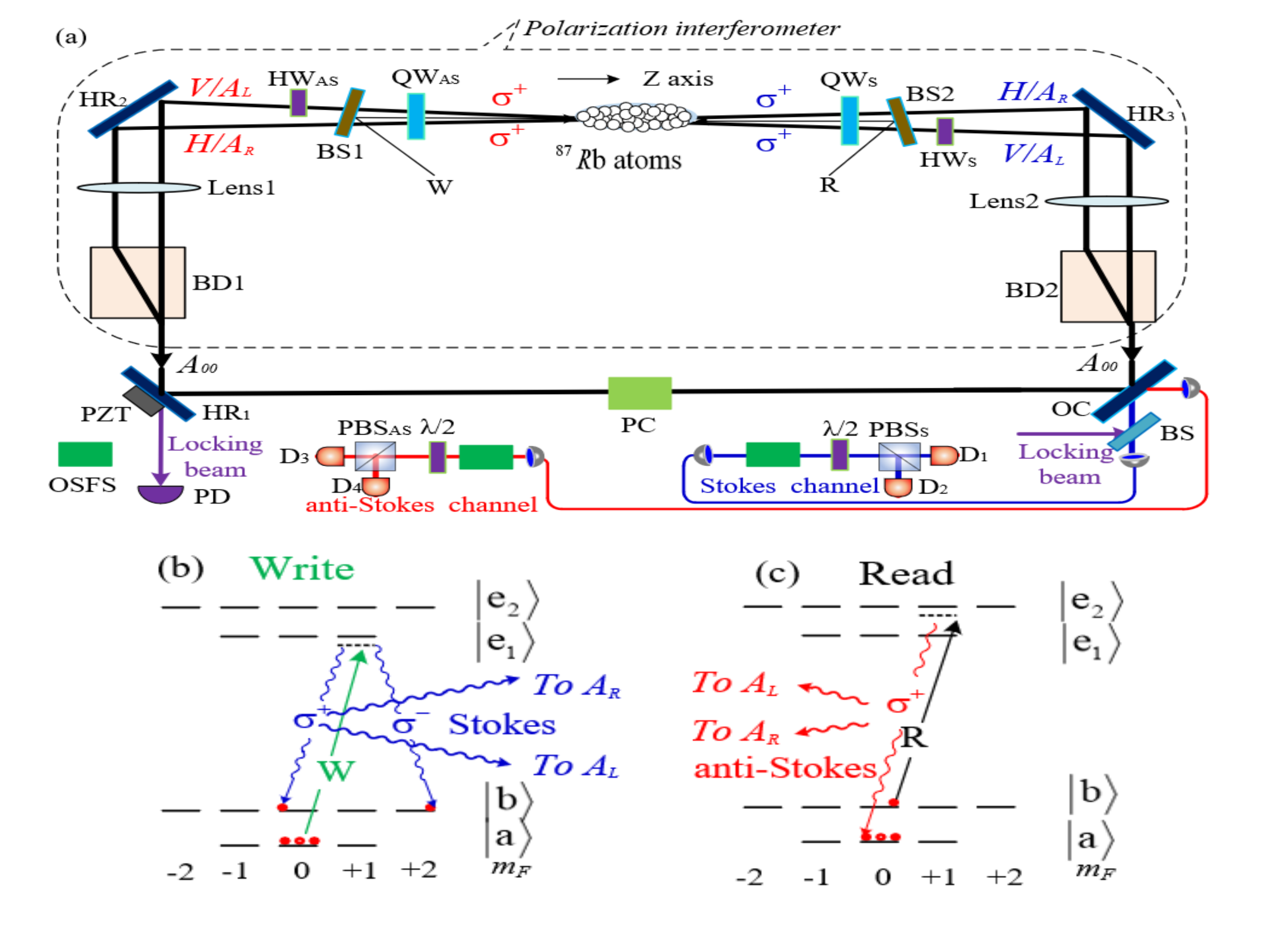}
     	\caption{ Schematic diagram for experimental setup (a): A polarization interferometer formed by two beam displacers (BD1 and BD2) is inserted into a ring cavity. A locking laser pulse is coupled to the A$_{00}$ mode through a beam splitter (BS). Leaks of the cavity-locking pulse from HR$_1$ are detected by a fast photodiode (PD) to generate error signals. The error signals are amplified and used to drive a piezoelectric transducer (PZT) to stabilize the cavity length. OSFS: optical-spectrum-filter set \cite{34}; PC: phase compensator; BS1 (BS2): non-polarizing beam splitter with a reflectance of 1\% (3\%). (b) and (c) are relevant atomic levels involved in the write and read processes, respectively, W (R): write (read) laser. }.
     	
     	\label{figure1}
     \end{figure}

       After a storage time t, we apply a ${\sigma ^{\rm{ + }}}$ -polarized read pulse onto the atoms through a beam splitter BS2. The read pulse is red-detuned by 110 MHz to the $\left| {\rm{b}} \right\rangle  \to \left| {{e_{\rm{2}}}} \right\rangle $ transition and counter-propagates with the write beam, which convert the spin-wave $\left| {{M_R}} \right\rangle $ ($\left| {{M_L}} \right\rangle $) into anti-Stokes photon. The anti-Stokes photon retrieved from  $\left| {{M_R}} \right\rangle $ ($\left| {{M_L}} \right\rangle $) is ${\sigma ^{\rm{ + }}}$ -polarized and emitted into the spatial mode determined by the wave-vector constraint $k_{A{S_R}}^{} \approx  - k_{{S_R}}^{}$ ($k_{A{S_L}}^{} \approx  - k_{{S_L}}^{}$), i.e., it propagates along anti-clockwise in the arm A$_R$ (A$_L$). The anti-Stokes fields in A$_R$ and A$_L$  are transformed into H-polarized fields by a $\lambda$/4-plate QW$_{AS}$. Furthermore, the anti-Stokes field in A$_L$ is transformed into V-polarized field by a $\lambda$/2-plate HW$_{AS}$. Both fields are combined into an anti-Stokes qubit $A{S_{qbit}}$ by BD1 and then propagate in A$_{00}$ mode along anti-clockwise. The $A{S_{qbit}}$  is split into H- and V- polarized fields by BD2, which direct into A$_R$ and A$_L$ modes, respectively. Next, both anti-Stokes modes are transformed into  ${\sigma ^{\rm{ + }}}$-polarization by wave-plates \cite{71} and then interact with the atoms. So, the atoms are repeatedly coupled with the cavity modes. The two-photon entangled state is written as:   
       \begin{equation}
       	\Phi _{{\rm{p - p}}}^{} 
       	= \left( {\left| H \right\rangle _S^{}\left| H \right\rangle _{AS}^{} + {e^{i({\varphi _S} + {\varphi _{AS}})}}\left| V \right\rangle _S^{}\left| V \right\rangle _{AS}^{}} \right)/\sqrt 2
       \end{equation}
       
       where, the subscript S (AS) denotes the Stokes (anti-Stokes) photon,${\varphi _{AS}}$  the relative phase between the two anti-Stokes emissions from the atoms to BD1 via A$_R$ and A$_L$ paths. In our experiment, the sum of ${\varphi _{AS}}$ and ${\varphi _{S}}$ is passively set to zero with a phase compensator \cite{71}. The cavity length is actively stabilized by coupling a cavity-locking beam through OC. The Stokes and anti-Stokes fields are tuned to resonate with the ring cavity \cite{71}. As shown in Fig.1, the escaped Stokes (anti-Stokes) photon from OC is coupled to a sing-mode fiber and then is guided into a polarization-beam splitter ${\rm{PBS}}_S^{}$ (${\rm{PBS}}_{AS}^{}$). Two outputs of ${\rm{PBS}}_S^{}$ (${\rm{PBS}}_{AS}^{}$) are sent to single-photon detectors ${D_{\rm{1}}}$ (${D_{\rm{3}}}$) and ${D_{\rm{2}}}$ (${D_{\rm{4}}}$). The polarization angle ${\theta _{S}}$ (${\theta _{AS}}$) of the Stokes (anti-Stokes) field is changed by rotating a $\lambda$/2-plate before${\rm{PBS}}_S^{}$ (${\rm{PBS}}_{AS}^{}$).	
    
	   \section{Results}
	    The intrinsic retrieval efficiency refers to spin-wave-to-photon conversion efficiency \cite{1, 17, 20, 24, 30, 33}, which is defined as the probability of converting the spin-wave excitation into the read-out photon in the a well-defined spatial-temporal mode (cavity mode) at the output of ensembles. The intrinsic retrieval efficiency of the SW qubit can be measured as 
	    $R_{qbi{\rm{t}}}^{inc} = P_{S,{\rm{ A}}S}^{}/\left( {{\eta _{TD}}P_S^{}} \right)$ \cite{71}, where $P_{S,{\rm{ A}}S}^{} = P_{{D_{\rm{1}}},{\rm{ }}{D_{\rm{3}}}}^{} + P_{{D_{\rm{2}}},{\rm{ }}{D_{\rm{4}}}}^{}$  denotes the probability of detecting a coincidence between the Stokes and anti-Stokes fields;$P_{{D_{\rm{1}}},{\rm{ }}{D_{\rm{3}}}}^{}$  ($P_{{D_{\rm{2}}},{\rm{ }}{D_{\rm{4}}}}^{}$ ) is the probability of detecting a coincidence between the detectors D$_1$ (D$_2$) and D$_3$  (D$_4$) for ${\theta _S} = {\theta _{AS}} = {\rm{0}}^\circ $  ,$P_S^{} = P_{{D_{\rm{1}}}}^{} + P_{{D_{\rm{2}}}}^{}$ the probability of detecting Stokes photon, $P_{{D_{\rm{1}}}}^{}$  ($P_{{D_{\rm{2}}}}^{}$ ) is the probability of detecting a photon at D$_1$ (D$_2$); ${\eta _{TD}} = {\eta _{esp}}{\eta _t}{\eta _D}$  is the total detection efficiency of the read-out (anti-Stokes) channel, which includes the efficiency of light escaping from the ring cavity,${\eta _{esp}} \approx 60\% $ , the transmission efficiency from the cavity to the detectors,${\eta _t} \approx 36\% $   \cite{71}, and the detection efficiency of the single-photon detectors,${\eta _D} \approx {\rm{68\% }}$  .Thus, the total detection efficiency is ${\eta _{TD}} \approx 15\% $  .Moreover, the intrinsic retrieval efficiency for an individual SW M$_R$ and M$_L$ mode is defined as $R_L^{inc} = P_{{D_{\rm{1}}},{\rm{ }}{D_{\rm{3}}}}^{}/\left( {{\eta _{TD}}P_{{D_{\rm{1}}}}^{}} \right)$   ($R_R^{inc} = P_{{D_{\rm{2}}},{\rm{ }}{D_{\rm{4}}}}^{}/\left( {{\eta _{TD}}P_{{D_{\rm{2}}}}^{}} \right)$).Fig.2 plots the measured efficiencies $R_{qbit}^{inc}$  (red circle dots), $R_{L}^{inc}$ (blue square dots), and $R_{R}^{inc}$   (green triangle dots) as functions of storage time t.From the figure, we see that $R_{qbit}^{inc} \approx R_L^{inc} \approx R_R^{inc}$  for different times t, which means that the retrieval efficiency for an SW qubit is the same as that for a single-mode SW. This shows that the efficiency loss of retrieving the qubit, which is a key limit in state-of-the-art work \cite{33}, is overcome in our experiment.The solid grey curve is the fit to the retrieval efficiencies $R_{qbit}^{inc}$ ,$R_{L}^{inc}$  and$R_{R}^{inc}$   according to the function $R(t){\rm{ = }}{R_{\rm{0}}}\left( {\exp \left( {{{{\rm{ - }}{{\rm{t}}^2}} \mathord{\left/
	    				{\vphantom {{{\rm{ - }}{{\rm{t}}^2}} {\tau _0^2}}} \right.
	    				\kern-\nulldelimiterspace} {\tau _0^2}}} \right){\rm{ + }}\exp \left( {{{{\rm{ - t}}} \mathord{\left/
	    				{\vphantom {{{\rm{ - t}}} {{\tau _0}}}} \right.
	    				\kern-\nulldelimiterspace} {{\tau _0}}}} \right)} \right)/2$ , which yields retrieval efficiencies${R_{\rm{0}}} = {\rm{77}}\% {\rm{  }}$  ,$R(t = 0.{\rm{23ms}}) \approx {\rm{66}}{\rm{.7\% }}$   and$R(t = {\rm{0}}.{\rm{54ms}}) \approx {\rm{50\% }}$  , together with a memory lifetime ${\tau _{\rm{0}}} \approx {\rm{1}}ms$ .In our presented memory, the SWs is stored in the “clock” coherence, thus, the influence on the memory lifetime of the inhomogeneous-broadening-induced decoherence is very small and then can be neglected (for details see Supplementary note 5 in Ref \cite{34}). The lifetime limited by SW decoherence due to the atomic random motions is estimated to be 1.4 ms \cite{71}. Such lifetime estimation is basically in agreement with our experimental result in Fig.2, which shows that our presented memory is mainly limited to the atomic-motion-induced decoherence.  
    			\begin{figure}[h]
    				\centering
    				\includegraphics[width=3in]{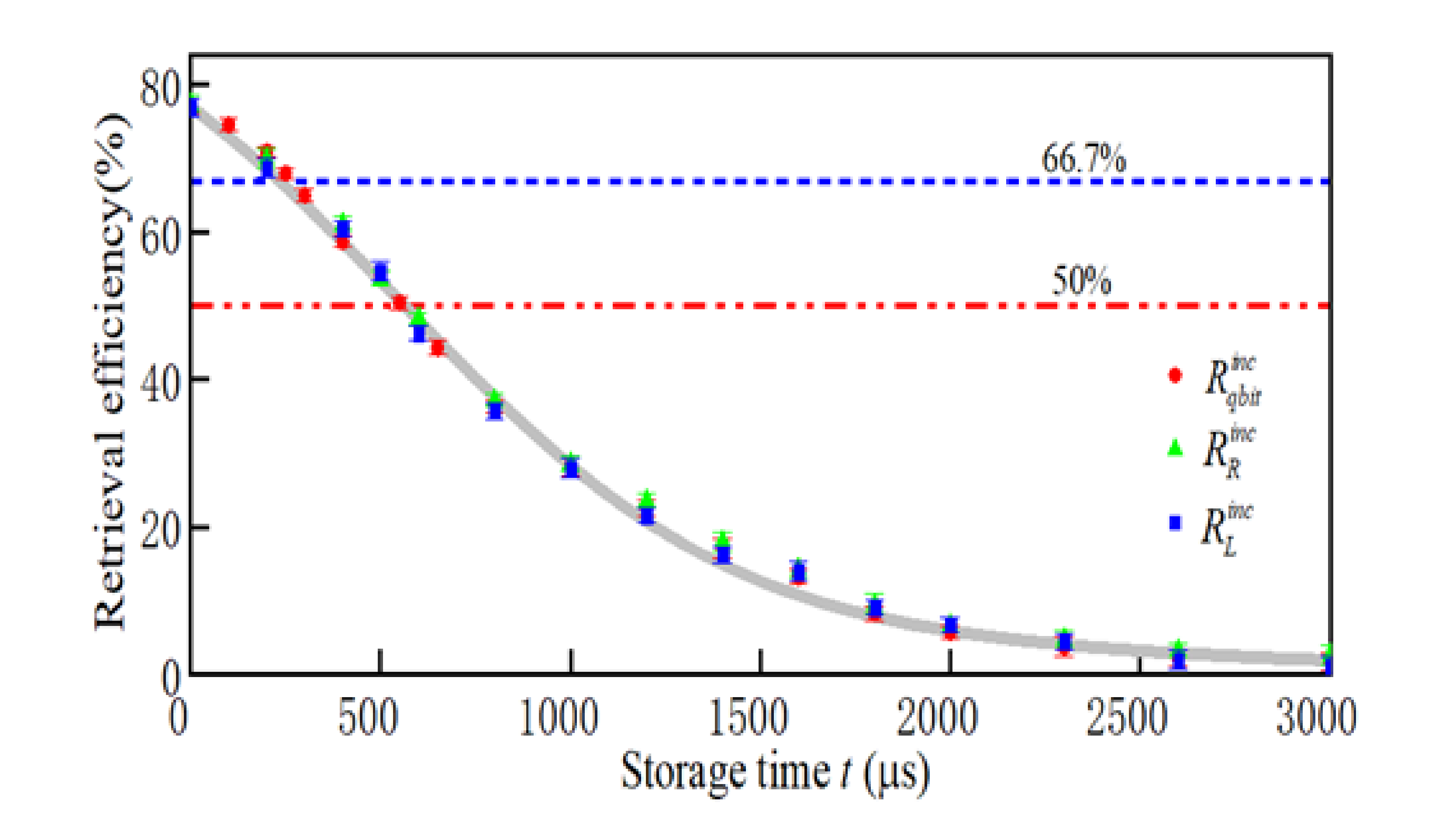}
    				\caption{ Intrinsic retrieval efficiencies as a function of storage time t for $\chi  = 1\% $ . Error bars represents 1 standard deviation.}.
    				
    				\label{fig:figure1}
    			\end{figure}	
	    
	    Next, we measure the Clauser–Horne–Shimony–Holt (CHSH) inequality, which is a type of Bell inequality, to confirm the spin-wave-photon entanglement state $\Phi _{{\rm{a - p}}}^{}$ .The Bell CHSH parameter is defined as ${S_{Bell}} = \left| \begin{array}{l}
	    	E({\theta _S},{\theta _{AS}}) - E({\theta _S},{{\theta '}_{AS}})\\
	    	+ E({{\theta '}_S},{\theta _{AS}}) + E({{\theta '}_S},{{\theta '}_{AS}})
	    \end{array} \right| < {\rm{2}}$  with the correlation function $E({\theta _S},{\theta _{AS}})$ , which is written as  $\frac{{{C_{{\rm{13}}}}({\theta _S},{\theta _{AS}}) + {C_{{\rm{24}}}}({\theta _S},{\theta _{AS}}) - {C_{{\rm{14}}}}({\theta _S},{\theta _{AS}}) - {C_{{\rm{23}}}}({\theta _S},{\theta _{AS}})}}{{{C_{{\rm{13}}}}({\theta _S},{\theta _{AS}}) + {C_{{\rm{24}}}}({\theta _S},{\theta _{AS}}) + {C_{{\rm{14}}}}({\theta _S},{\theta _{AS}}) + {C_{{\rm{23}}}}({\theta _S},{\theta _{AS}})}}$. For example,  ${C_{{\rm{13}}}}({\theta _S},{\theta _{AS}})$ (${C_{{\rm{24}}}}({\theta _S},{\theta _{AS}})$ ) denotes the coincidence counts between the detectors  D$_1$ (D$_2$) and D$_3$  (D$_4$) for the polarization angles  ${\theta _S}$ and  ${\theta _{AS}}$.We used the canonical settings ${\theta _S} = {\rm{0}}^\circ $ ,${\theta '_S} = {\rm{45}}^\circ $  ,${\theta _{AS}} = {\rm{22}}{\rm{.5}}^\circ $  and ${\theta '_{AS}} = {\rm{67}}{\rm{.5}}^\circ $  in measuring the Bell parameter${S_{Bell}}$  .Fig. 3 shows the decay of ${S_{Bell}}$  as a function of storage time t (blue squares) for $\chi  = 2\% $ .At$t \approx {\rm{0 }}\mu {\rm{s}}$  ,${S_{Bell}} = {\rm{2}}{\rm{.5}} \pm {\rm{0}}{\rm{.02}}$  , while at t =1.15 ms, ${S_{Bell}} = {\rm{2}}{\rm{.05}} \pm {\rm{0}}{\rm{.03}}$ . These violate the Bell inequality by 25 and 1.7 standard deviations, respectively.Furthermore, at t=2.6 ms,${S_{Bell}} = {\rm{1}}{\rm{.15}} \pm {\rm{0}}{\rm{.03}}$   . From Fig.3, one can see that S parameter degrades with storage time t. The reason for this degradation mainly results from SW decaying with t \cite{1, 72, 73}, which has been observed in Fig.2. 
	    
	    	\begin{figure}[h]
	    	\centering
	    	\includegraphics[width=3in]{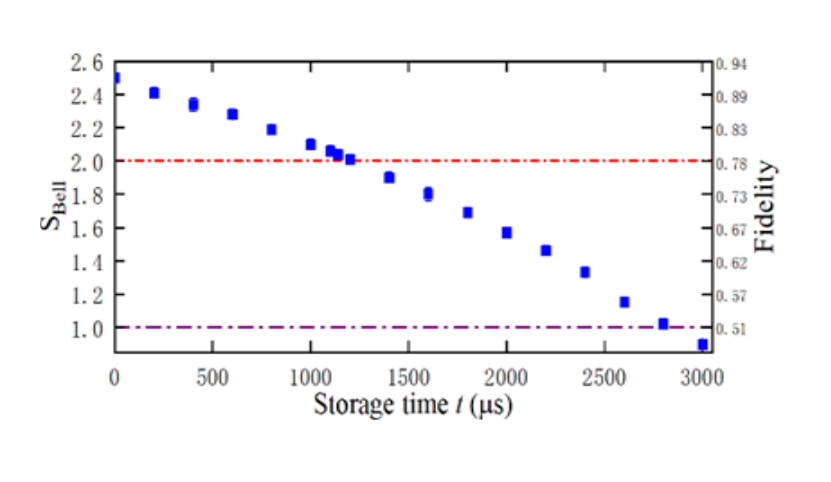}
	    	\caption{ Measured Bell parameter as a function of storage time t for $\chi  = 2\% $. Error bars represent 1 standard deviation.}.
	    	
	    	\label{fig:figure1}
	    \end{figure}
    
	With the measured Bell parameters, we may estimate the fidelity F of the atom-photon entangled state. For an entangled state, the relationship of visibility V and Bell parameter S can be expressed as $ V = S/2\sqrt 2$ \cite{74} and the relationship of the entanglement fidelity F and the visibility V can be expressed as $ F = \left( {{\rm{3V + 1}}} \right)/4 $ \cite{75}. So, the relationship of F and S can be written as $F{\rm{ = }}\left( {{\rm{3}}S/2\sqrt 2 {\rm{ + 1}}} \right)/4$  . According to this relationship, we obtain that the measured S=1.15 corresponds to a F=0.55, which exceeds the bound of 0.5 required to observe entanglement for a Bell state.

    At zero delay, the measured data of S parameter in our experiment is higher than the measured result of 2.30(3) in Ref.\cite{31}, but lower than the measured result of 2.64(9) in Ref.\cite{33}. Compared to the experiment in Ref.\cite{33}, where the angles are $>2.5^0$, the angles in our experiment are set to be very small ($0.053^0$) for achieving long wavelength storage. While, the small-angle storage leads to a bad separation between write (read) beam and write-out (read-out) detection channel, which then degrade the quality of the entangled state. When using optical-lattice atoms instead of MOT cold atoms, we may give up the small-angle storage scheme used for suppressing the atomic-motion-induced decoherence.
     
	For comparison, we plot Fig. 4 to show the measured retrieval efficiencies for the memories restricted to qubit storages as functions of storage times in various systems via different schemes. One can see that our experimental result of 50\% (66.7\%) intrinsic retrieval efficiency for a storage time of 540 µs (230 µs) is 13.5 (24) times higher than the previously reported best result \cite{33} (\cite{30}). 
		\begin{figure}[h]
		\centering
		\includegraphics[width=3in]{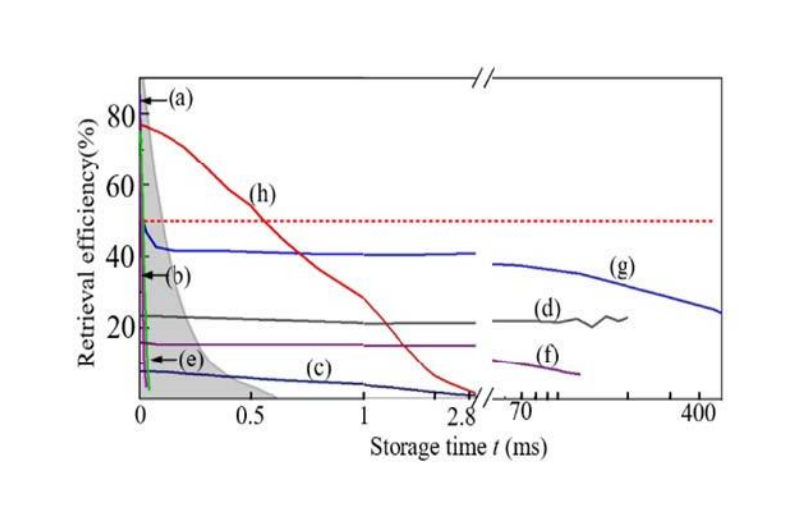}
		\caption{Retrieval efficiencies of memory qubit vs storage time t in various systems. The shaded region shows the result for an ideal-optical-fiber loop. Curves (a) and (b) are the results for storages of single-photon \cite{50} and weak-coherent-light \cite{49} qubits via EIT in high-optical-density cold atoms. Curve (c) represents the results for qubit memory stored as MFI spin waves via EIT in cold atoms \cite{44}. Curve (d) represents single-photon qubit storage via EIT in single atoms \cite{10}. Curves (e), (f), (g) and (h), corresponding to the results in Refs.\cite{30}, \cite{32}, \cite{33} and our experiment, respectively, represent qubit memories via DLCZ protocol.}.
		
		\label{fig:figure1}
	\end{figure}

	    \section{Conclusion}
	     The use of the polarization interferometer in the cavity enables us to apply a write beam to create two SW modes of the memory qubit, both associated with the clock coherence and coupled to the ring cavity. Thanks to phase-matching condition, the two SW modes are retrieved by a read beam. The relative phase between the two arms is passively stabilized and easily set to be zero by the phase compensator.On this basis, we achieve quadruple resonance of the cavity with the Stokes and retrieved fields propagating in the cavity modes A$_R$ and A$_L$, respectively. This represents the first demonstration of an atomic ensemble simultaneously coupled into two TEM$_0$$_0$ modes of a cavity.
	     Due to the uses of the two-arm ring cavity, our presented setup avoids the double-mode retrieval efficiency loss in Ref. \cite{33}. We achieve atom-photon entanglement with zero-delay retrieval efficiency of 77\% and millisecond lifetime.At 50\% retrieval efficiency, the storage time reaches $540\mu s$, which is 13.5 times longer than the best reported result \cite{33}. 
	     To generate and verify heralded entanglement in a repeater link with long separation distances requires to use long-lifetime and high-efficiency atom-photon entanglement interface as repeater nodes. For example, for generating heralded entanglement in a L=100 km link, the storage time of the nodes is required to exceed the link communication time of L/c =500$\mu$s  , where $c=2*10^8m/s$ is the light speed in fibers. Combining quantum frequency conversion that change Stokes photon from 795 nm into the telecommunications band \cite{9,31,35,76}, our presented setup can be well used for demonstrating such a repeater link. More important, at this storage time, our presented setup still has 50\% retrieval efficiency, which promises one to effectively verify the heralded entanglement. 
	     In the quantum memories for a single mode SW using optical-lattice atoms\cite{24, 33}, the storage time at 50\% retrieval efficiency reaches 50 milliseconds, which is far longer than that in the memories for qubit. In our presented experiment, the storage lifetime is limited to SW decoherence caused by motions of cold atoms. If we use optical-lattice atoms instead of the cold atoms in the presented setup, the storage time at 50\% qubit retrieval efficiency will be extended to 50 milliseconds, which would be of benefit in QR-based long-distance quantum communications.    
	     Furthermore, by selecting more cavity modes to couple with the atoms, we will achieve massively-multiplexed and high-reversible spin-wave–photon entanglement with long lifetime and then is used for achieving long-distance (1000-km) entanglement distributions through QR \cite{71}.
	     
	     \part*{{\tiny {\normalsize {\large {\Large Acknowledgements}}}}}
	     
	     We acknowledge funding support from Key Project of the Ministry of Science and Technology of China (Grant No. 2016YFA0301402); The National Natural Science Foundation of China (Grants: No. 11475109, No. 11974228), Fund for Shanxi “1331 Project” Key Subjects Construction.

		\bibliographystyle{quantum}
		
     	\onecolumn
    	\appendix
	
	   \section{Supplementary material}
	   \subsection{Two-mode ring cavity based on polarization interferometer}
	   As shown in the Fig.1 in the main text, our experimental setup is mainly formed by the ring optical cavity inserted by the polarization interferometer with cold atoms in it. As mentioned in the main text, the ring cavity is formed by three flat mirrors (HR$_{1,2,3}$) and a flat output coupler (OC). The cold atoms are placed at location O. To explain our experimental scheme, we firstly discuss the ring cavity without the polarization interferometer, which is shown in Fig.5. 
	   \begin{figure}[h]
	   	\centering
	   	\includegraphics[width=3in]{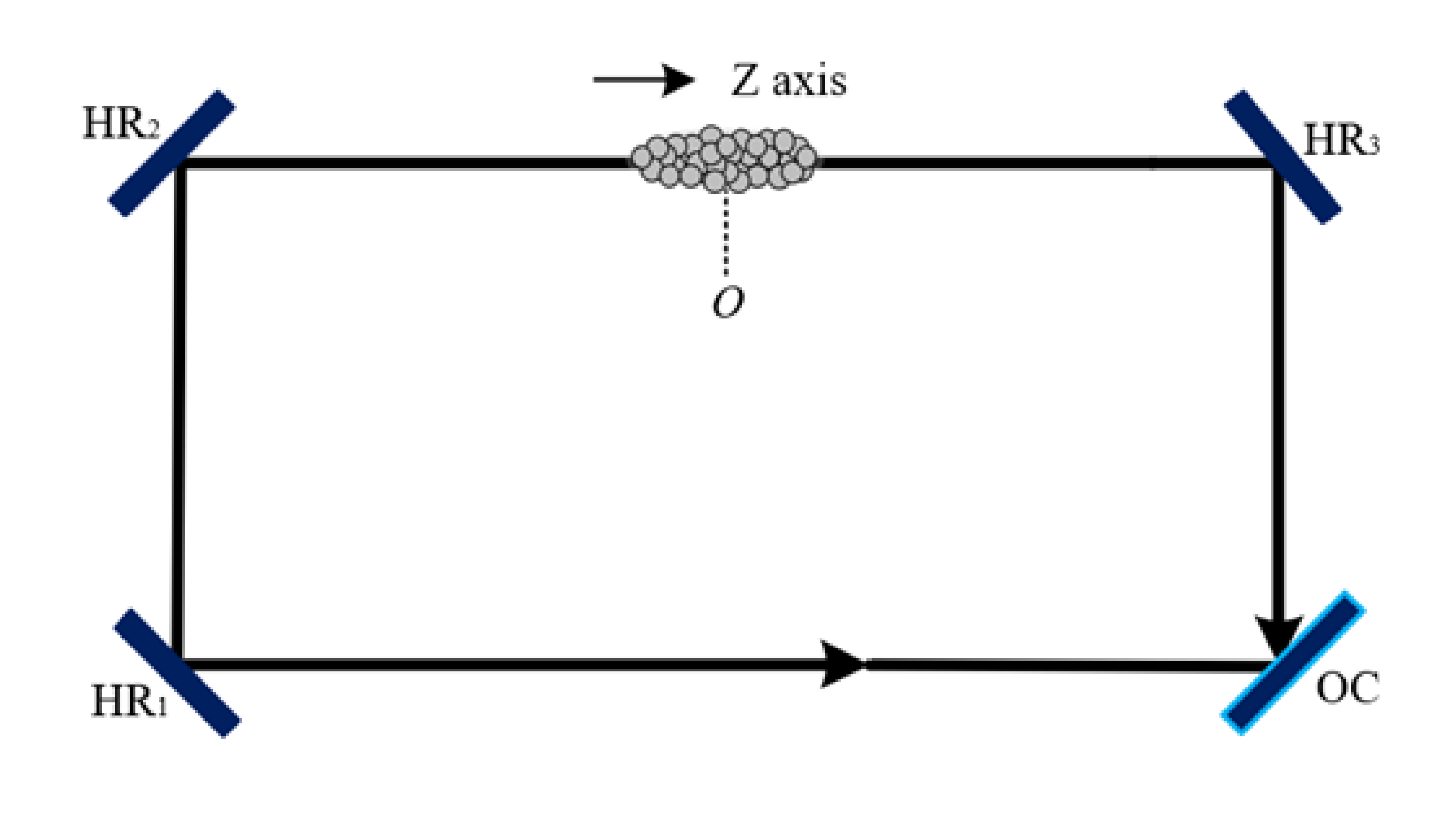}
	   	\caption{Ring cavity without the polarization interferometer}.
	   	
	   	\label{fig:figure1}
	   \end{figure}
	    The transfer (ABCD) matrix of the ring cavity with a length of L is given by ${T_{C1}} = \left[ {\begin{array}{*{20}{c}}
	    		A&B\\
	    		C&D
	    \end{array}} \right] = \left[ {\begin{array}{*{20}{c}}
	    		{\rm{1}}&L\\
	    		{\rm{0}}&{\rm{1}}
	    \end{array}} \right]$ , which show that the ring cavity satisfy the critical condition $(A + D)/2 = 1$ . By aligning the cavity mirrors, we may make a TEM$_{00}$ light beam, which has a small divergence angle, bounce back and forth in the cavity. The propagating path of the cavity mode A$_{00}$denotes as the black line in Fig.5. To decrease the escaping loss of the mode A$_{00}$ when it propagates in the cavity, we set its spot size (diameter) to be large, whose value is 5.2 mm in our presented experiment. The measured loss of the mode A$_{00}$ escaping from the cavity per round trip is 2.7 \%, which is far less than the 20\% transmission of OC and allows the mode A$_{00}$ to serve as the cavity mode. As shown in Fig. 5, the A$_{00}$ cavity mode goes though the cold atoms, which is coincided with the z-axis during the path from HR$_{2}$ to HR$_{3}$. In our experiment, the z-axis is quantum direction defined by a bias magnetic field (4G), along which the write beam is applied. 
        \begin{figure}[h]
        	\centering
        	\includegraphics[width=3in]{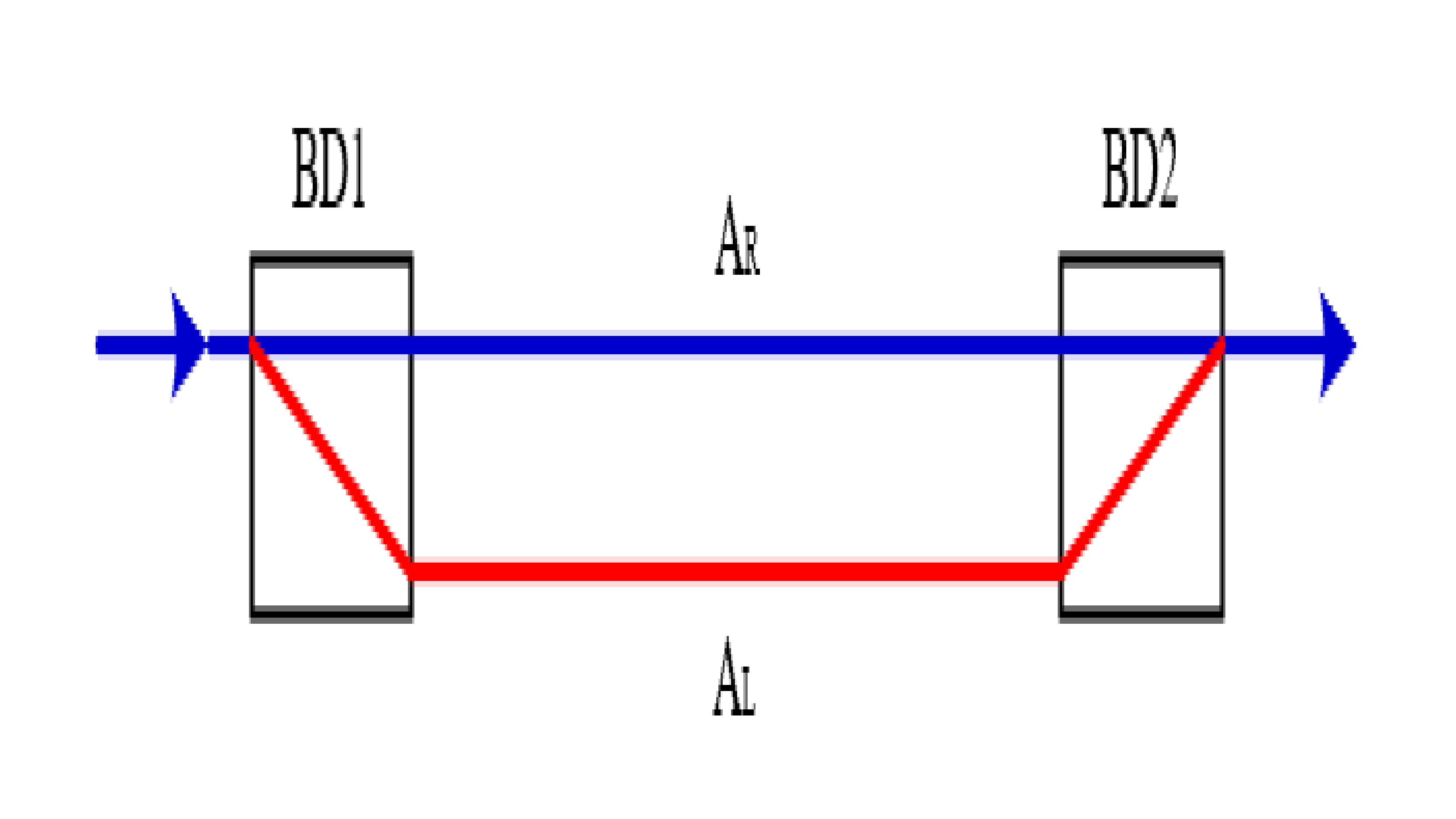}
        	\caption{A common polarization interferometer}.
        	
        	\label{fig:figure1}
        \end{figure}

        A common polarization interferometer is formed by two identical beam displacers (BD1 and BD2) separated by a distance. Fig.6 shows the configuration of the common polarization interferometer, where, an arbitrarily-polarized light beam moves towards the right and enters the BD1. The BD1 split the light beam into H-polarized and V-polarized components. Both beam components direct into A$_{R}$ and A$_{L}$ arms of the interferometer, respectively, parallel propagate, and then are recombined into a light beam by BD2. 
        
        \begin{figure}[h]
        	\centering
        	\includegraphics[width=4in]{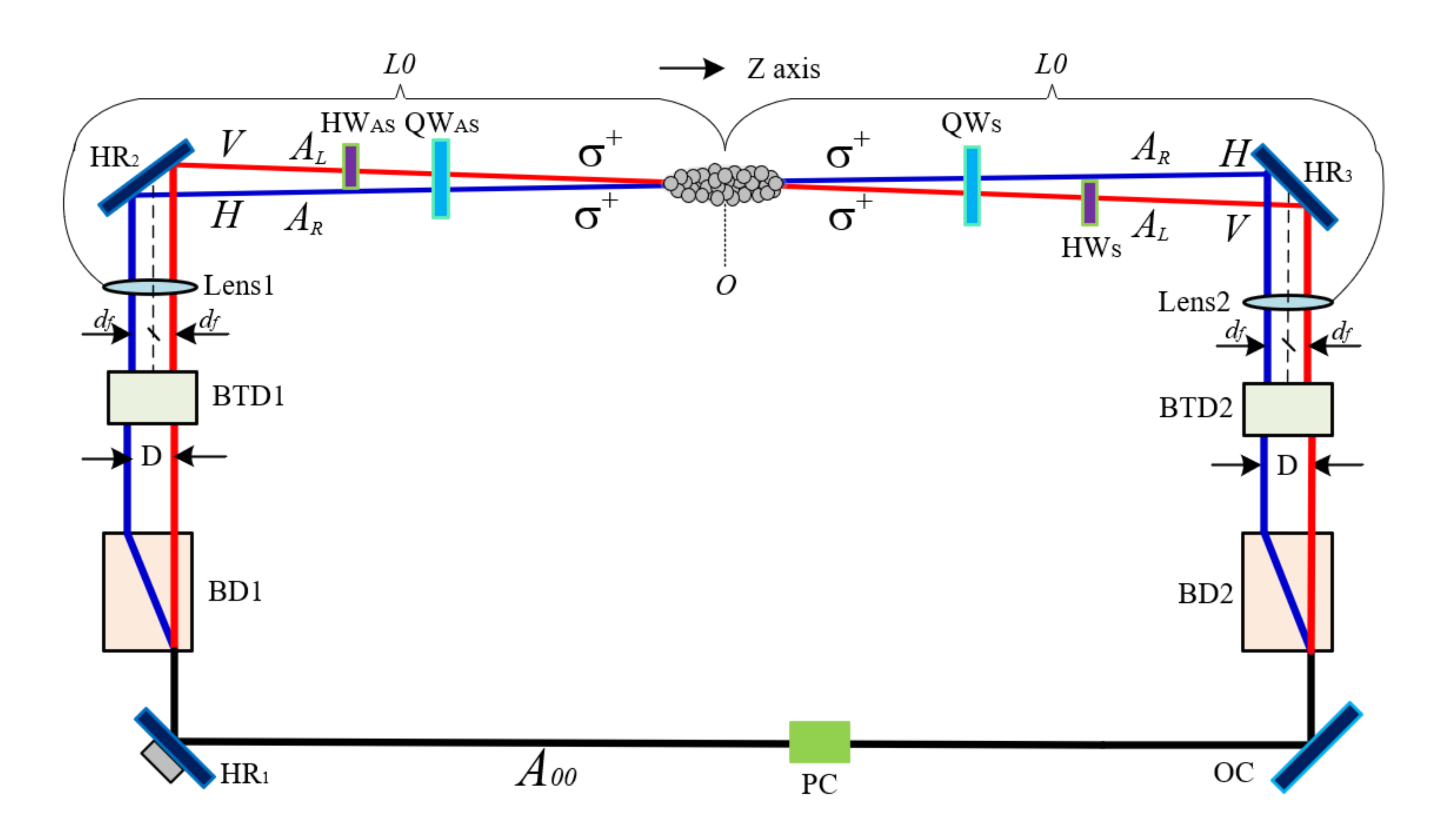}
        	\caption{ Diagram of the two-mode ring cavity based on the polarization interferometer}.
        	
        	\label{fig:figure1}
        \end{figure}
        In our experimental set-up, the polarization interferometer is combined into the ring cavity and the two arms are required to couple with the cold atomic ensemble. For which, BD1 (BD2) is inserted in between HR$_1$ and HR$_2$ (HR$_3$ and OC) and two lenses (lens1 and lens2) are inserted in between two beam displacers.As shown in Fig.7, the lens1 (lens2) is inserted in between BD1 and HR$_{2}$ (between BD2 and HR$_{3}$).For setting the polarizations of the two arms to be a specific polarization (${\sigma ^{\rm{ + }}}$ -polarization), we insert the 1/4 wave-plates QW$_A$$_S$ and QW$_S$ in the two arms and 1/2 wave-plates QW$_A$$_S$ and QW$_S$ only in A$_L$ arm. The optical axis of BD1 (BD2) is arranged to be parallel with the A$_0$$_0$ path line in the cavity. The lengths from the lens1 or lens2 to the atomic center O is L$_{0}$. Both lenses have the same focal length F$_0$= L$_0$. We also insert two beam-transformation devices, i.e., BTD1 and BTD2, in between BD1 and lens1 and BD2 and lens2, respectively. BTD1 (BTD2) is formed by two lenses, which can shrink (expand) the two-separation distance between the two arms by a factor of F$_{BTD} $ when they propagate along clockwise [1]. Importantly, the two arms remain parallel after going through BTD. We now explain the paths of the two arms of the interferometer and then discuss the reason for introducing the BTDs. When A$_{00}$ mode propagates along clockwise in the cavity, it will be split into H- and V- polarized components by BD1. The H-polarized and V-polarized components of the A$_0$$_0$ mode propagate in the two arms (A$_R$ and A$_L$), respectively, whose separation distance denotes by D. After going through BTD1, the arms A$_R$ and A$_L$ still propagate parallel but their distance reduces to D/F$_
        {BTD}$. As shown in Fig.7, we carefully arrange the lens1 to make the distance from its main optical axis (dashed line) to the A$_R$ or A$_L$ path line be all equal to d$_{f}$. Both components of the A$_0$$_0$ mode are reflected by HR$_{2}$. The V-polarized component is transformed into H-polarization by the 1/2-wave-plate HW$_{AS}$. Subsequently, both H-polarized components are transformed into ${\sigma ^{\rm{ + }}}$ -polarization by the 1/4-wave-plate QW$_{AS}$ and then crossways go through the atoms (the location O). The angle between the arm A$_R$ (A$_L$) and the write beam (along z-axis) is given by ${\theta _R} = {d_f}/{L_0}$   ( ${\theta _L} = {d_f}/{L_0}$  ).Next, both ${\sigma ^{\rm{ + }}}$  -polarized components are transformed into H-polarization by the 1/4-wave-plate QW$_S$. The H-polarized component in A$_L$ arm is transformed into V-polarization by the 1/2-wave-plate HW$_S$. After reflected by HR$_3$, both components go through the lens2. We also arrange the lens2 to make the distances from its main optical axis to the A$_R$ or A$_L$ arm be all equal to d$_{f}$. After the lens2, the two arms parallel propagate, whose separation distance is D/F$_{BTD}$, which is the same as that before the lens1. The two arms go through BTD2, which expand their separation distance by the factor of F$_{BTD}$. Thus, after BTD2, the separation distance of the two arms is transformed back into D. The two arms parallel propagate and enter the BD2, which combine the two arms into the A$_0$$_0$ mode. In our presented experiment, the distance between lens1 and lens2 is carefully set to be 2F$_{0}$ in order to enable the combination of the two arms via BD2 is perfect. The measured loss of the mode A$_{00}$ via A$_R$ (A$_L$) path escaping from cavity per round trip is 3.2\% (3.2\%), showing that the interferometer-based cavity supports the A$_R$ and A$_L$ arms to serve as cavity modes. 
        
        In our experiment, we have to individually operate polarizations of the Stokes (anti-Stokes) photons propagating in the arms A$_R$ and A$_L$. For example, we insert the 1/2 wave-plate HW$_{AS}$ (HW$_S$) only in the A$_L$ arm after BD1(BD2), which transform its H-polarization to V-polarization. For achieving the individual polarization operations, the separation distance D between the two arms is required to be more than the diameter of the cavity mode A$_0$$_0$. In our presented experiment, we set D=5.5 mm, which is larger than the diameter of A$_0$$_0$. Without the BTDs, the separation distance of the two arms at lens1 (lens2) will be D=2d$_{f}$ =5.5mm. To achieve a long-lived memory, we have to suppress decoherence due to atomic motions, which in turn require to achieve long wave-length SW storage and then decrease the angle  [1]. With BTD1, the separation distance 2df at the lens1 (lens2) is changed into 2d$_{f}$=D/F$_{BTD}$. To achieve small values of the angles, we use the lenses with F0 =1.5m and BTD1 with factor of F$_{BTD}$=2. So, the angles can be given as ${\theta _R} = {\theta _L} = {d_f}/\left( {{F_{BTD}}{L_0}} \right) = {d_f}/\left( {{F_{BTD}}{F_0}} \right) \approx {0.053^0}$  , which corresponds to a millisecond lifetime [1]. It is noted that for briefly and clearly explaining the ring cavity combined with the polarization interferometer, we neglected the BTDs in the Fig.1 in the main text.
        
        In our presented experiment, the measured total cavity loss is 13\%, which includes: (1) Transmission losses due to the non-polarization beam splitters BS1 and BS2 (see Fig.1 in the main text), which are 1\% and 3\%, respectively; (2) Optical loss due to imperfect reflections of HR$_{1,2,3,}$ which is 1\%; (3) Transmission losses of the optical elements (BDs, BTDs, lens1,2, and wave-plates, et.al.), which is 4.8\%; (4) Loss of the mode A$_R$ (A$_L$ ) escaping from the cavity per round trip, which is 3.2\% (3.2\%). The length of the ring cavity is 6 m, corresponding to a free spectral range of 50 MHz. The refection rate of the OC is 80\%. The cavity has finesses of 16.9 and 17.0 for the modes A$_R$ and A$_L$. The spot sizes of both modes at the location O (the center of the atoms) are all 0.5mm.
        
        The dispersion of the Stokes and anti-Stokes fields due to the beam displacers is very small and then can be neglected. In our experiment, the relative phase between two Stokes (anti-Stokes) fields propagating from BD1 to BD2 via A$_R$ and A$_L$ arms is compensated to be zero with the phase compensator (labeled as PC in Fig.1). Thus, the sum of ${\varphi _S}$  and ${\varphi _{AS}}$  is set to zero. Since the phase compensator is formed by wave-plates [1], this phase compensation is passively. By changing the frequency of the write (read) laser beam, we tune the frequency of the Stokes (anti-Stokes) field and then make Stokes (anti-Stokes) field resonate with the cavity. The Stokes emissions into A$_R$ and A$_L$ modes are both enhanced by a factor of ${\raise0.7ex\hbox{${2F}$} \!\mathord{\left/
        		{\vphantom {{2F} \pi }}\right.\kern-\nulldelimiterspace}
        	\!\lower0.7ex\hbox{$\pi $}}$ by the cavity. Also, the efficiencies for retrieving the two SWs M$_R$ and M$_L$ are enhanced through the Purcell effect [2].
        
         \subsection{Experimental details}
         
         The experiment is performed in a cyclic fashion. In each experimental cycle, the durations for preparing cold atoms and running the experiment for spin-wave-photon entanglement (SWPE) generation are 42 ms and 8 ms, respectively, corresponding to a 20-Hz cycle frequency. During the preparation stage, more than $10^8$ atoms of $^{87}$Rb are trapped in a two-dimensional magneto-optical trap (MOT) for 41.5 ms and further cooled via Sisyphus cooling for 0.5 ms.The cloud of cold atoms has a size of 5×2×2 mm$^3$, a temperature of 100 $\mu$K, and an optical density of 16. At the end of this preparation stage, a bias magnetic field of B$_0$=4G is applied along the z-axis (see Fig. 1a), and the atoms are optically pumped into the initial level $\left| {{5^2}{S_{{1 \mathord{\left/
         					{\vphantom {1 2}} \right.
         					\kern-\nulldelimiterspace} 2}}},F = 1,m = 0} \right\rangle $ . After the preparation stage, the 8-ms experimental run containing a large number of SWPE-generation trials starts. At the beginning of a trial, a write pulse with a duration of 300 ns is applied to the atomic ensemble to generate correlated pairs of Stokes photons and spin-wave excitations. The detection events at the Stokes detectors D$_1$  and D$_2$    in Fig. 1a are analyzed with a field-programmable gate array (FPGA).As soon as a Stokes photon qubit is detected by either one of these detectors, SWPE is generated and the FPGA sends out a feed-forward signal to stop the write processes. After a storage time t, a read laser pulse with a duration of 300 ns is applied to the atoms to convert the spin-wave qubit into the anti-Stokes photon qubit. After a 1300-ns interval, a cleaning pulse with a duration of 200 ns is applied to pump the atoms into the initial level $\left| {{5^2}{S_{{1 \mathord{\left/
         					{\vphantom {1 2}} \right.
         					\kern-\nulldelimiterspace} 2}}},F = 1,m = 0} \right\rangle $ .  Then, the next SWPE-generation trial starts. However, in most cases, the Stokes photon is not detected during the write pulse owing to the low excitation probability ($\chi  \le 2\% $ ). If this is the case, the atoms are pumped directly back into the initial level by the read and cleaning pulses. Subsequently, the next trial starts, i.e., the write pulse is applied. The delay between the two adjacent write pulses for a storage time of $t \approx {\rm{1 \mu s}}$  is 2000 ns.Therefore, the 8-ms experimental run contains 4000 experimental trials. Considering that a 1-s experiment contains 20 cycles, the repetition rate of the SWPE-generation trail is $r = 8 \times {10^4}$  .
         				
         	The powers of the two beams applied onto the atoms are 300 $\mu$W and 10 mW, respectively. The read laser is red-detuned by 110 MHz to the transition $\left| {\rm{b}} \right\rangle  \to \left| {{{\rm{e}}_2}} \right\rangle $ .  To block the write (read) laser beam in the Stokes and anti-Stokes channels, we place an optical-spectrum-filter set (OSFS) before each polarization beam splitter PBS$_S$ (PBS$_{AS}$).Each OSFS comprises five Fabry–Perot etalons, which attenuate the write (read) beam by a factor of $8.1*10^{-12}$ ($3.8*10^{-11}$ ) and transmit the Stokes (anti-Stokes) fields with a transmission of ~56\%. Additionally, in the Stokes (anti-Stokes) detection channel, the spatial separation of the Stokes (anti-Stokes fields) from the strong write (read) beam attenuates the write (read) beam by a factor of $10^{-4}$ .In this experiment, we measured the uncorrelated noise probability in the anti-Stokes mode, which is ${p_N} \approx {10^{ - 4}}$  per read pulse (300 ns). Such uncorrelated noise mainly results from the leakage of the read beam into the anti-Stokes detection channel. 
         	
         	The efficiency of the Stokes (anti- Stokes) photons escaping from the ring cavity is defined as  ${\eta _{esp}} = \frac{{{T_{OC}}}}{{{T_{OC}} + L}}$, where ${T_{OC}} = 20\% $   is the transmission of the output coupler mirror, and $L \approx 13\% $   is the cavity total loss. Thus, we have ${\eta _{esp}} \approx 60.6\% $ .
         	
         	The transmission efficiency  ${\eta _T}$ of the Stokes (anti-Stokes) photons from the cavity to the detectors includes the coupling efficiency ${\eta _{SMF}} \approx {\rm{0}}{\rm{.71}}$  of the single-mode fiber SMF$_{AS}$, the transmission ${\eta _{Filter}} \approx {\rm{0}}{\rm{.56}}$  of the optical-spectrum-filter set (OSFS), and the transmission  ${\eta _{MMF}} \approx {\rm{0}}{\rm{.92}}$ of the multi-mode fiber. Therefore, ${\eta _T} = {\rm{0}}{\rm{.71}} \times {\rm{0}}{\rm{.56}} \times {\rm{0}}{\rm{.92}} \approx 36.6\% $.
         	
         	The total efficiency of detecting the Stokes (anti-Stokes) photon is given by ${\eta _{TD}} = {\eta _{esp}} \times {\eta _T} \times {\eta _D}$  , where  ${\eta _D}$ is the quantum efficiency of the single-photon detectors ( D$_1$ , D$_2$ , D$_3$ and D$_4$), whose value is  ${\eta _D} \approx 68\%$ in our experiment. So, we have ${\eta _{TD}} = {\eta _{esp}} \times {\eta _T} \times {\eta _D} \approx 15\% $ . 
         	
         	All error bars in the experimental data represent a ±1 standard deviation, which is estimated from Poissonian detection statistics using Monte Carlo simulations.
         	
         	 \subsection{Measurement for the intrinsic retrieval efficiency }
         In DLCZ memory experiments, the relationship of the probabilities of detecting a Stokes, anti-Stokes photons and coincidence can written as[3]: \[{P_{S,}}_{aS} = \chi {R^{inc}}{\eta _S}{\eta _{aS}} + {P_S}{P_{aS}}\]
         
         where, $\chi $ is the excitation probability,  ${P_{S,}}_{aS}$ the probability of detecting a coincidence between the Stokes and anti-Stokes fields,  ${P_S} = \chi {\eta _S} + B{\eta _S}$ (${P_{aS}} = \chi {R^{inc}}{\eta _{aS}} + C{\eta _{aS}}$ ) the probability of detecting photon in the Stokes (anti-Stokes) field,  ${\eta _{aS}}$ and  ${\eta _{S}}$  are the overall detection efficiencies in the anti-Stokes and Stokes channels, respectively, B (C) is the background noise in the Stokes (anti-Stokes) channel.From this relationship, the intrinsic retrieval efficiency can be given by  ${R^{inc}} = \frac{{{P_{S,}}_{aS} - {P_S}{P_{aS}}}}{{\chi {\eta _S}{\eta _{aS}}}} = \frac{{{P_{S,}}_{aS} - {P_S}{P_{aS}}}}{{\left( {{P_S} - B{\eta _S}} \right){\eta _{aS}}}}$ . It also can be written as ${R^{inc}} = \frac{{{R^{net}}}}{{{\eta _{aS}}}}$ , where, ${R^{net}} = \frac{{{P_{S,}}_{aS} - {P_S}{P_{aS}}}}{{{P_S} - B{\eta _S}}}$  denotes the net retrieval efficiency, which has been used in Ref.[4].For the case that the background noise B is very small and can be neglected, one has ${R^{inc}} = \frac{{{P_{S,}}_{aS} - {P_S}{P_{aS}}}}{{{P_S}{\eta _{aS}}}} = \frac{{{P_{S/}}_{aS} - {P_{aS}}}}{{{\eta _{aS}}}}$ , which has been used in Ref. [5], where, ${P_{aS/S}}$  refers to the detection probability of a anti-Stokes photon conditioned upon the detection of a photon in the Stokes field.  In the case of that the excitation probability $\chi $   and the background noise B and C are all very small, the accident coincidence  ${P_S}{P_{aS}}$ will be far less than the coincidence probability  ${P_{S,}}_{aS}$  and then ${R^{inc}} \approx \frac{{{P_{S,}}_{aS}}}{{{\eta _{aS}}{P_S}}}$ . In our presented work, the excitation probability  $\chi  \sim {\rm{0}}{\rm{.01/trial}}$ and the background noise $B \sim {\rm{1}}{{\rm{0}}^{ - 5}}$  and $C \sim 5 \times {\rm{1}}{{\rm{0}}^{ - 4}}$ are all very small, we then measure the intrinsic retrieval efficiency of the SW qubit according to $R_{qbi{\rm{t}}}^{inc} = P_{S,{\rm{ A}}S}^{}/\left( {{\eta _{TD}}P_S^{}} \right)$ , where $P_{S,{\rm{ A}}S}^{} = P_{{D_{\rm{1}}},{\rm{ }}{D_{\rm{3}}}}^{} + P_{{D_{\rm{2}}},{\rm{ }}{D_{\rm{4}}}}^{}$ ,$P_{{D_{\rm{1}}},{\rm{ }}{D_{\rm{3}}}}^{}$   ($P_{{D_{\rm{2}}},{\rm{ }}{D_{\rm{4}}}}^{}$ ) is the probability of detecting a coincidence between the detectors  D$_1$   (  D$_2$) and  D$_3$ ( D$_4$ ) for ${\theta _S} = {\theta _{AS}} = {\rm{0}}^\circ $ ,$P_S^{} = P_{{D_{\rm{1}}}}^{} + P_{{D_{\rm{2}}}}^{}$  , $P_{{D_{\rm{1}}}}^{}$ ($P_{{D_{\rm{2}}}}^{}$) is the probability of detecting a photon at D$_1$   (D$_2$).
         
          \subsection{Estimation of the memory lifetime limited by atomic-motion-induced decoherence}
         The lifetime ${\tau _a}$  limited by atomic-motion-induced decoherence can be expressed by ${\tau _a} = \left( {{k_w} - {k_S}} \right)\sin \theta /{v_a}$ [1, 6], where,${v_a} = \sqrt {{k_B}T/m} $   denotes the average atomic speed, ${k_B}$ Boltzmann’s constant, T the average temperature of the atoms,${k_w}$   (${k_S}$  ) denotes the wave-vector of the write beam (Stokes photon),$\theta  = \left| {{\theta _R}} \right| = \left| {{\theta _L}} \right|$.According to this expression and our experimental set up with $\theta  = {\rm{0}}{\rm{.05}}{{\rm{3}}^{\rm{0}}}$ , we estimated that the memory lifetime ${\tau _a}$  to be ~1.4 ms. 
          \subsection{Improvement on the total detection efficiencies}
          
          The low total detection efficiency can be effectively improved in future work. Assuming that the cavity loss reduces to 1\%, the cavity escaping efficiency ${\eta _{esp}}$  will be up to 95\%.urthermore, the efficiencies ${\eta _{SM{F_S}}}$ , ${\eta _{Filter}}$ , and ${\eta _{MMF}}$  can be greatly improve to ~99\%, ~99\%, and ~99\%.One can also use superconductor single-photon detectors with a detection efficiency of 95\% [7] instead of silicon-avalanche-photodiode single-photon detectors. With the aforementioned improvements, the total detection efficiency would change to \[{\eta _{TD}} = {\eta _{esp}} \times {\eta _{SM{F_S}}} \times {\eta _{Filter}} \times {\eta _{MMF}} \times {\eta _D} \approx 0.98*0.99*0.98*0.99*0.95 \approx 88\% \]
           \subsection{Repeater rates for distributing an entangled photon pair over distances up to 1000 km through a multiplexed QR using cavity-perfectly-enhanced or cavity-imperfectly-enhanced schemes}
          On the basis of two-photon interference, the repeater rate for distributing an entangled photon pair over a distance   through multiplexed [8-11] DLCZ QR protocols with nest level n can be evaluated by the following equation [12]:
          \[{R_{rate}} \approx \frac{1}{{{T_{cc}}}}P_0^{(N)}\left( {\prod\nolimits_{j = 1}^{j = 4} {{P_j}} } \right){P_{pr}}\]
          where ${T_{cc}} = {L_0}/c$ is the communication time with c being the speed of light in fibers and ${L_0} = L/n$  the distance between two nodes belonging to an elementary link;  $P_0^{(N)} = 1 - {(1 - {P_0})^N} \approx N{P_0}$ is the success probability for entanglement generation in an individual elementary link using multiplexed nodes that each store N qubits,with ${P_0} = \left( {{\chi ^2}{e^{ - {L_0}/{L_{att}}}}\eta _{FC}^2\eta _{TD}^2} \right)/2$ being that using non-multiplexed nodes, $\eta _{FC}^{}$  being the memory-to-telecom frequency conversion efficiency, $\eta _{TD}^{}$  being the total detection efficiency, and the factor 1/2 being due to double-excitation events from a single node [12, 13]; ${P_j} = \left[ {{{\left( {{R_0}{e^{ - {t_{j - 1}}/{\tau _0}}}} \right)}^2}\eta _{TD}^2} \right]/2$ with $j = 1{\rm{ to }}n = 4$  is the success probability for entanglement swapping at the i-th level, with ${\tau _0}$  the lifetime of the multiplexed node memory based on an atomic ensemble, and   the qubit retrieval efficiency of the memory at zero delay;${t_0} \simeq {T_{cc}}/P_0^{(N)} = {T_{cc}}/\left( {N{P_0}} \right)$  is the time needed for the elementary entanglement generation; ${t_j} \simeq {t_{j - 1}}/{P_j}$ $j = 1{\rm{ to }}n = 4$  is the time needed for the ith-level entanglement swapping; and ${P_{pr}} \approx {\left( {{R_0}{e^{ - {t_4}/{\tau _0}}}} \right)^2}/2$  is the probability for distributing an entangled photon pair over the distance L. 
          
          In an optical EIT storage [14], the 1/e lifetime of ${\tau _0} = 16{\rm{ s}}$  by apply dynamic decoupling pulse sequences has been achieved. Quantum memories with comparable storage time may be within reach. The multiplexed storages of qubits with hundreds of spatial modes [15] and tens of temporal modes [16] have been demonstrated via DLCZ protocol in cold atoms. Thus, one can achieve multiplexed qubit memories with thousands of modes by combining both schemes [17, 18] into individual systems. For the case that high-performance quantum technologies including large-scale-multiplexing (e.g., 1000 modes), cavity-enhanced-retrieval and long-lifetime (16 s) quantum storages, high-efficiency memory-to-telecom frequency conversion (with efficiency of 33\%) [13] as well as superconductor single-photon detections [7] (with efficiency of 95\%) are integrated into individual repeater nodes, we calculate the repeater rate for distributing an entangled photon pair over distance L through 4-level QR through Eq.(S1) and present the results in Fig.8.The blue solid and red dashed curves are the results for the cases of using cavity-perfectly-enhanced (CPE) retrieval ( ${R_0} \approx 80\% $ ) and cavity-imperfectly-enhanced (CIE) retrieval (${R_0} \approx 60\% $) schemes, respectively, where, the other parameters are the same.For a fixed repeater rate of $10^{-4}$, one can see that the QR using CPE-retrieval nodes may achieve an entangled photon pair distribution over $L \approx 1000km$  , while that using CIE retrieval nodes may only achieve entanglement distribution over  $L \approx 430km$  , showing a significant advance of CPE over CIE retrieval schemes. 
        \begin{figure}[h]
        	\centering
        	\includegraphics[width=5in]{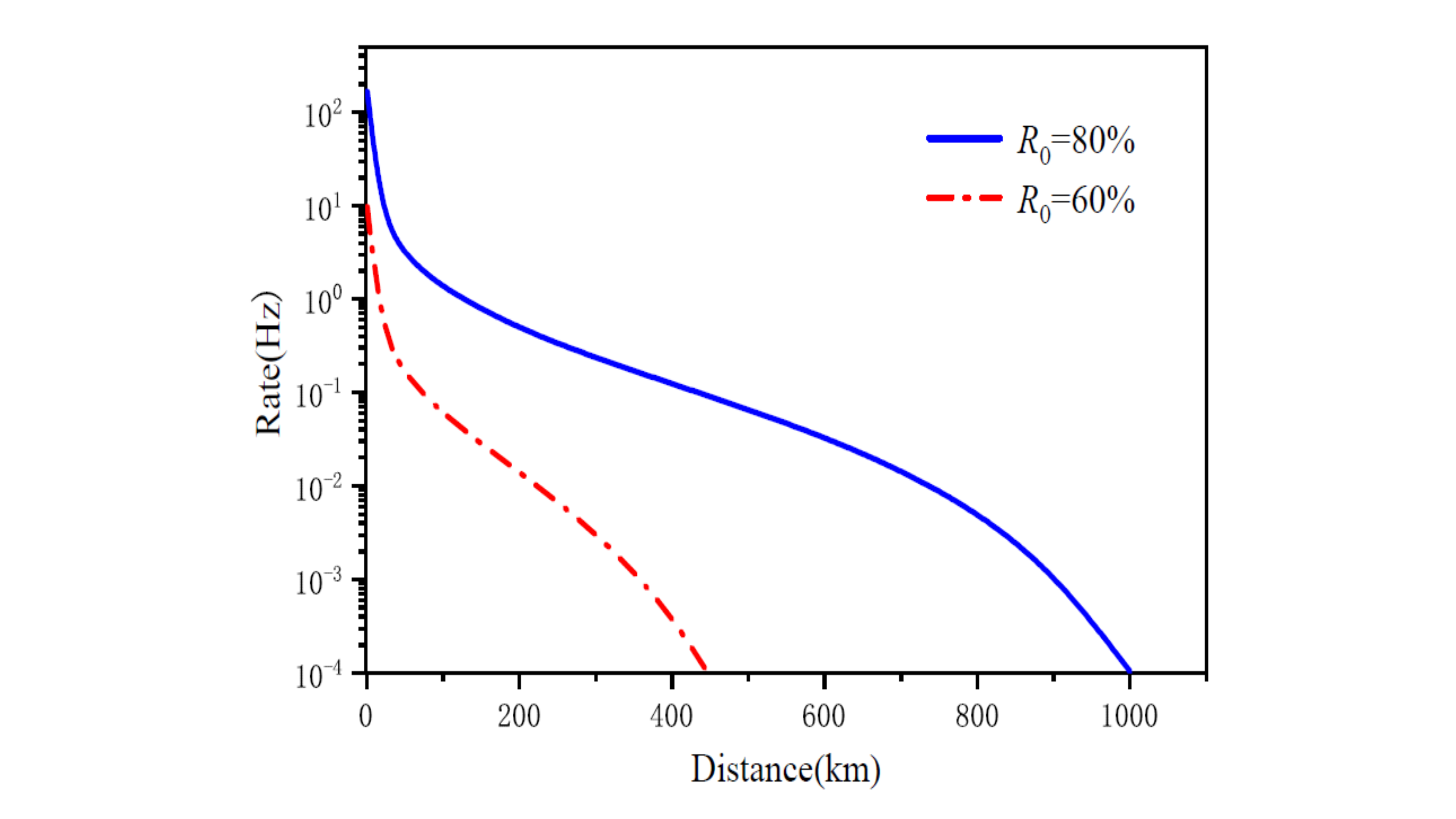}
        	\caption{Calculated repeater rates as functions of entanglement distribution over distance L according to Eq. S1, where the blue solid (red dash) curve corresponds to the quantum repeater using nodes with zero-delay retrieval efficiency ${R_0} \approx 80\% $   (${R_0} \approx 60\% $  ). The parameters used in the calculation are: nest level n=4, mode number N=1000, memory lifetime ${\tau _0} = 16{\rm{ s}}$  , total detection efficiency ${\eta _{TD}} \approx 88\% $  for the Stokes (anti-Stokes) detection channel, and quantum frequency conversion efficiency ${\eta _{FC}} = 33\% $ .}.
        	
        	\label{fig:figure1}
        \end{figure}
      \bibliographystyle{quantum}

\begin{thebibliography}{9}
		
			\bibitem{1}
			N. Sangouard, C. Simon, H. de Riedmatten, N. Gisin, Quantum repeaters based on atomic ensembles and linear optics.
			\href{https://doi.org/10.1103/RevModPhys.83.33}{Rev. Mod. Phys. \textbf{83}, 33-80 (2011).}


	       \bibitem{2}
            L. M. Duan, M. D. Lukin, J. I. Cirac, a. P. Zoller, Long-distance quantum communication with atomic ensembles and linear optics. 
           \href{ https://doi.org/10.1038/35106500}{Nature \textbf{414},  413-418 (2001).}

                 
            \bibitem{3}
            C. Simon, Towards a global quantum network.  
           \href{ https://doi.org/10.1038/s41566-017-0032-0}{Nat. Photon. \textbf{11},   678-680 (2017).}
           
           
           \bibitem{4}
           H. J. Kimble, The quantum internet.  
           \href{   https://doi.org/10.1038/nature07127    }{   Nature       \textbf{ 453  },     1023-1030  (2008) .}


            \bibitem{5}
           S. Wehner, D. Elkouss, R. Hanson, Quantum internet: A vision for the road ahead. 
           \href{    https://doi.org/10.1126/science.aam9288   }{     Science      \textbf{  362  }, 303 (2018). }
           
           
            \bibitem{6}
            P. Kómár, E. M. Kessler, M. Bishof, L. Jiang, A. S. Sørensen, J. Ye, M. D. Lukin, A quantum network of clocks. 
           \href{   https://doi.org/10.1038/nphys3000    }{     Nat. Phys.    \textbf{  10  },  582-587 (2014).      }
           
           
            \bibitem{7}
           F. Bussières, N. Sangouard, M. Afzelius, H. de Riedmatten, C. Simon, W. Tittel, Prospective applications of optical quantum memories.   
           \href{     https://doi.org/10.1080/09500340.2013.856482  }{      Journal of Modern Optics   \textbf{  60   },   1519-1537 (2013).    }
           
           
            \bibitem{8}
            A. I. Lvovsky, B. C. Sanders, W. Tittel, Optical quantum memory. 
           \href{ https://doi.org/10.1038/nphoton.2009.231 }{Nat. Photon.   \textbf{ 3   }, 706-714 (2009).  }
           
           
            \bibitem{9}
           T. van Leent, M. Bock, R. Garthoff, K. Redeker, W. Zhang, T. Bauer, W. Rosenfeld, C. Becher, H. Weinfurter, Long-Distance Distribution of Atom-Photon Entanglement at Telecom Wavelength.  
           \href{https://doi.org/10.1103/PhysRevLett.124.010510    }{Phys. Rev. Lett.  \textbf{124    },010510 (2020).   }
           
           
           \bibitem{10}
            M. Körber, O. Morin, S. Langenfeld, A. Neuzner, S. Ritter, G. Rempe, Decoherence-protected memory for a single-photon qubit.   
           \href{  https://doi.org/10.1038/s41566-017-0050-y }{ Nat. Photon.   \textbf{12},18-21 (2017).  }
           
           
           \bibitem{11}
           D. L. M. B. B. Blinov, L.-M Duan, C. Monroe	Observation of entanglement between a single trapped atom and a single photon. 
           \href{ https://doi.org/10.1038/nature02377      }{Nature          \textbf{ 428   },153-157(2004).        }
           
           
           \bibitem{12}
           B. Hensen, H. Bernien, A. E. Dreau, A. Reiserer, N. Kalb, M. S. Blok, J. Ruitenberg, R. F. Vermeulen, R. N. Schouten, C. Abellan, W. Amaya, V. Pruneri, M. W. Mitchell, M. Markham, D. J. Twitchen, D. Elkouss, S. Wehner, T. H. Taminiau, R. Hanson, Loophole-free Bell inequality violation using electron spins separated by 1.3 kilometres.  
           \href{https://doi.org/10.1038/nature15759  }{ Nature    \textbf{ 526 }, 682-686 (2015).  }
           
           
           \bibitem{13}
           A. Delteil, Z. Sun, W.-b. Gao, E. Togan, S. Faelt, A. Imamoğlu, Generation of heralded entanglement between distant hole spins.   
           \href{https://doi.org/10.1038/nphys3605  }{Nat. Phys. \textbf{12    },218-223 (2015). }
           
           
           \bibitem{14}
           N. G. Mikael Afzelius, and Hugues de Riedmatten, Quantum memory for photons.   
           \href{https://doi.org/10.1063/PT.3.3021       }{ Physics Today        \textbf{ 68   }, 12, 42 (2015).   }
           
           
           \bibitem{15}
           K. Heshami, D. G. England, P. C. Humphreys, P. J. Bustard, V. M. Acosta, J. Nunn, B. J. Sussman, Quantum memories: emerging applications and recent advances.  
           \href{  https://doi.org/10.1080/09500340.2016.1148212 }{Journal of Modern Optics \textbf{63    },2005-2028 (2016).  }
           
           
           \bibitem{16}
            A. Kuzmich, W. P. Bowen, A. D. Boozer, A. Boca, C. W. Chou, L.-M. Duan, a. H. J. Kimble, Generation of nonclassical photon pairs for scalable quantum communication with atomic ensembles. 
           \href{ https://doi.org/10.1038/nature01714   }{ Nature  \textbf{ 423  }, 731-734 ( 2003).  }
           
           
           \bibitem{17}
           Julien Laurat, Hugues de Riedmatten, Daniel Felinto, Chin-Wen Chou, E. W., Schomburg, a. H. J. Kimble., Schomburg, and H. Jeff Kimble. Efficient retrieval of a single excitation stored in an atomic ensemble.  
           \href{  https://doi.org/10.1364/OE.14.006912  }{   Opt. Express  \textbf{ 14  }, 6912-6918 (2006).  }
           
           
           \bibitem{18}
           J. Simon, H. Tanji, J. K. Thompson, V. Vuletic, Interfacing collective atomic excitations and single photons.   
           \href{ https://doi.org/10.1103/PhysRevLett.98.183601   }{  Phys. Rev. Lett.  \textbf{98   }, 183601 (2007). }
           
           
           \bibitem{19}
           D. Felinto, C. W. Chou, H. de Riedmatten, S. V. Polyakov, H. J. Kimble, Control of decoherence in the generation of photon pairs from atomic ensembles.  
           \href{https://doi.org/10.1103/PhysRevA.72.053809  }{  Phys. Rev. A  \textbf{72   },053809 (2005). }
           
           
           \bibitem{20}
            B. Zhao, Y.-A. Chen, X.-H. Bao, T. Strassel, C.-S. Chuu, X.-M. Jin, J. Schmiedmayer, Z.-S. Yuan, S. Chen, J.-W. Pan, A millisecond quantum memory for scalable quantum networks.  
           \href{https://doi.org/10.1038/nphys1153      }{ Nat. Phys. \textbf{5   }, 95-99 (2008).    }
           
           
           \bibitem{21}
           X.-H. Bao, A. Reingruber, P. Dietrich, J. Rui, A. Dück, T. Strassel, L. Li, N.-L. Liu, B. Zhao, J.-W. Pan, Efficient and long-lived quantum memory with cold atoms inside a ring cavity. 
           \href{https://doi.org/10.1038/nphys2324  }{  Nat. Phys. \textbf{8 },517-521 (2012). }
           
           
           
           \bibitem{22}
            R. Zhao, Y. O. Dudin, S. D. Jenkins, C. J. Campbell, D. N. Matsukevich, T. A. B. Kennedy, A. Kuzmich, Long-lived quantum memory.    
           \href{ https://doi.org/10.1038/nphys1152   }{Nat. Phys. \textbf{5   }, 100-104 (2008).  }
           
           
           \bibitem{23}
           A. G. Radnaev, Y. O. Dudin, R. Zhao, H. H. Jen, S. D. Jenkins, A. Kuzmich, T. A. B. Kennedy, A quantum memory with telecom-wavelength conversion. 
           \href{https://doi.org/10.1038/nphys1773      }{ Nat. Phys.         \textbf{6   }, 894-899 (2010).     }
           
           
           \bibitem{24}
           S.-J. Yang, X.-J. Wang, X.-H. Bao, J.-W. Pan, An efficient quantum light–matter interface with sub-second lifetime.    
           \href{https://doi.org/10.1038/nphoton.2016.51      }{Nat. Photon.  \textbf{ 10  },381-384 (2016). }
           
           
           \bibitem{25}
           E. Bimbard, R. Boddeda, N. Vitrant, A. Grankin, V. Parigi, J. Stanojevic, A. Ourjoumtsev, P. Grangier, Homodyne tomography of a single photon retrieved on demand from a cavity-enhanced cold atom memory.  
           \href{https://doi.org/10.1103/PhysRevLett.112.033601      }{ Phys. Rev. Lett.  \textbf{ 112   }, 033601 (2014).     }
           
           
           \bibitem{26}
            H. Li, J.-P. Dou, X.-L. Pang, T.-H. Yang, C.-N. Zhang, Y. Chen, J.-M. Li, I. A. Walmsley, X.-M. Jin, Heralding quantum entanglement between two room-temperature atomic ensembles.   
           \href{https://doi.org/10.1364/optica.424599      }{ Optica        \textbf{8   }, 925-929 (2021).    }
           
           
            \bibitem{27}
           K. B. Dideriksen, R. Schmieg, M. Zugenmaier, E. S. Polzik, Room-temperature single-photon source with near-millisecond built-in memory. 
           \href{https://doi.org/10.1038/s41467-021-24033-8      }{  Nat. Commun.         \textbf{12   }, 3699 (2021).    }
           
           
            \bibitem{28}
           C. Laplane, P. Jobez, J. Etesse, N. Gisin, M. Afzelius, Multimode and Long-Lived Quantum Correlations Between Photons and Spins in a Crystal.  
           \href{  https://doi.org/10.1103/PhysRevLett.118.210501     }{ Phys. Rev. Lett.        \textbf{118   }, 210501 (2017).    }
           
           
            \bibitem{29}
            K. Kutluer, M. Mazzera, H. de Riedmatten, Solid-State Source of Nonclassical Photon Pairs with Embedded Multimode Quantum Memory. 
           \href{ https://doi.org/10.1103/PhysRevLett.118.210502  }{  Phys. Rev. Lett.   \textbf{118   }, 210502 (2017).  }
           
           
            \bibitem{30}
           S. J. Yang, X. J. Wang, J. Li, J. Rui, X. H. Bao, J. W. Pan, Highly retrievable spin-wave-photon entanglement source.  
           \href{https://doi.org/10.1103/PhysRevLett.114.210501      }{ Phys. Rev. Lett.       \textbf{114     }, 210501 (2015).     }
           
           
            \bibitem{31}
           Y. Yu, F. Ma, X. Y. Luo, B. Jing, P. F. Sun, R. Z. Fang, C. W. Yang, H. Liu, M. Y. Zheng, X. P. Xie, W. J. Zhang, L. X. You, Z. Wang, T. Y. Chen, Q. Zhang, X. H. Bao, J. W. Pan, Entanglement of two quantum memories via fibres over dozens of kilometres.  
           \href{ https://doi.org/10.1038/s41586-020-1976-7      }{Nature         \textbf{578   }, 240-245 (2020).     }
           
           
           \bibitem{32}
           Y. O. Dudin, A. G. Radnaev, R. Zhao, J. Z. Blumoff, T. A. Kennedy, A. Kuzmich, Entanglement of light-shift compensated atomic spin waves with telecom light. 
           \href{https://doi.org/10.1103/PhysRevLett.105.260502      }{  Phys. Rev. Lett.        \textbf{105   }, 260502 (2010).     }
           
           
           \bibitem{33}
           X. J. Wang, S. J. Yang, P. F. Sun, B. Jing, J. Li, M. T. Zhou, X. H. Bao, J. W. Pan, Cavity-Enhanced Atom-Photon Entanglement with Subsecond Lifetime. 
           \href{https://doi.org/10.1103/PhysRevLett.126.090501  }{  Phys. Rev. Lett.  \textbf{ 126  }, 090501 (2021). }
           
           
           \bibitem{34}
            S.-Z. Wang, M.-J. Wang, Y.-F. Wen, Z.-X. Xu, T.-F. Ma, S.-J. Li, H. Wang, Long-lived and multiplexed atom-photon entanglement interface with feed-forward-controlled readouts. 
           \href{https://doi.org/10.1038/s42005-021-00670-9      }{ Communications Physics         \textbf{ 4  }, 168 (2021).    }
           
           
           \bibitem{35}
            R. Ikuta, T. Kobayashi, T. Kawakami, S. Miki, M. Yabuno, T. Yamashita, H. Terai, M. Koashi, T. Mukai, T. Yamamoto, N. Imoto, Polarization insensitive frequency conversion for an atom-photon entanglement distribution via a telecom network.
           \href{  https://doi.org/10.1038/s41467-018-04338-x     }{ Nat. Commun. \textbf{9   }, 1997 (2018).     }
           
           
           \bibitem{36}
            Y. F. Pu, N. Jiang, W. Chang, H. X. Yang, C. Li, L. M. Duan, Experimental realization of a multiplexed quantum memory with 225 individually accessible memory cells.  
           \href{ https://doi.org/10.1038/ncomms15359     }{ Nat. Commun.        \textbf{ 8  }, 15359 (2017).    }
           
           
           \bibitem{37}
           Y.-F. Pu, S. Zhang, Y.-K. Wu, N. Jiang, W. Chang, C. Li, L.-M. Duan, Experimental demonstration of memory-enhanced scaling for entanglement connection of quantum repeater segments. 
           \href{ https://doi.org/10.1038/s41566-021-00764-4      }{  Nat. Photon.        \textbf{ 15  }, 374-378 (2021).     }
           
           
           \bibitem{38}
            H. Tanji, S. Ghosh, J. Simon, B. Bloom, V. Vuletic, Heralded single-magnon quantum memory for photon polarization States. 
           \href{ https://doi.org/10.1103/PhysRevLett.103.043601   }{Phys. Rev. Lett.   \textbf{103   }, 043601 (2009).   }
           
           
           \bibitem{39}
           Liu C, Dutton Z, Behroozi C H , H. L. V, Observation of coherent optical information storage in an atomic medium using halted light pulses.
           \href{ https://doi.org/10.1038/35054017  }{  Nature   \textbf{409   }, 490–493 (2001).    }
           
           
            \bibitem{40}
            D. F. Phillips, A. Fleischhauer, A. Mair, R. L. Walsworth, M. D. Lukin, Storage of light in atomic vapor. 
           \href{ https://doi.org/10.1103/PhysRevLett.86.783     }{ Phys. Rev. Lett.       \textbf{86   }, 783-786 (2001).       }
           
           
            \bibitem{41}
            M. D. Eisaman, A. Andre, F. Massou, M. Fleischhauer, A. S. Zibrov, M. D. Lukin, Electromagnetically induced transparency with tunable single-photon pulses. 
           \href{ https://doi.org/10.1038/nature04327     }{ Nature         \textbf{438   }, 837-841 (2005).     }
           
           \bibitem{42}
           T. Chaneliere, D. N. Matsukevich, S. D. Jenkins, S. Y. Lan, T. A. Kennedy, A. Kuzmich, Storage and retrieval of single photons transmitted between remote quantum memories. 
           \href{  https://doi.org/10.1038/nature04315     }{ Nature         \textbf{438   }, 833-836 (2005).    }
           
           
            \bibitem{43}
           K. S. Choi, H. Deng, J. Laurat, H. J. Kimble, Mapping photonic entanglement into and out of a quantum memory.   
           \href{ https://doi.org/10.1038/nature06670     }{  Nature        \textbf{ 452  }, 67 (2008).   }
           
           
            \bibitem{44}
           Z. Xu, Y. Wu, L. Tian, L. Chen, Z. Zhang, Z. Yan, S. Li, H. Wang, C. Xie, K. Peng, Long lifetime and high-fidelity quantum memory of photonic polarization qubit by lifting zeeman degeneracy.  
           \href{ https://doi.org/10.1103/PhysRevLett.111.240503     }{ Phys. Rev. Lett.        \textbf{ 111   }, 240503 (2013).    }
           
           
            \bibitem{45}
           S-Y Zhou, S-C Zhang, C. Liu, J. F. Chen, Jianming Wen, M. M. T. Loy, G. K. L.Wong, S.-W. Du, Optimal storage and retrieval of single-photon waveforms. 
           \href{ https://doi.org/10.1364/OE.20.024124     }{ Optics express         \textbf{20   }, 24124 (2012).     }
           
           
            \bibitem{46}
           G. Heinze, C. Hubrich, T. Halfmann, Stopped light and image storage by electromagnetically induced transparency up to the regime of one minute.
           \href{ https://doi.org/10.1103/PhysRevLett.111.033601      }{  Phys. Rev. Lett.        \textbf{ 111  }, 033601 (2013).    }
           
           
            \bibitem{47}
           Y. H. Chen, M. J. Lee, I. C. Wang, S. Du, Y. F. Chen, Y. C. Chen, I. A. Yu, Coherent optical memory with high storage efficiency and large fractional delay.  
           \href{https://doi.org/10.1103/PhysRevLett.110.083601      }{Phys. Rev. Lett.         \textbf{ 110  }, 083601 (2013).      }
                      
                   
           
            \bibitem{48}
           Y. F. Hsiao, P. J. Tsai, H. S. Chen, S. X. Lin, C. C. Hung, C. H. Lee, Y. H. Chen, Y. F. Chen, I. A. Yu, Y. C. Chen, Highly Efficient Coherent Optical Memory Based on Electromagnetically Induced Transparency.  
           \href{https://doi.org/10.1103/PhysRevLett.120.183602      }{ Phys. Rev. Lett.\textbf{120   }, 183602 (2018).}
           
           
           
            \bibitem{49}
            P. Vernaz-Gris, K. Huang, M. Cao, A. S. Sheremet, J. Laurat, Highly-efficient quantum memory for polarization qubits in a spatially-multiplexed cold atomic ensemble. 
           \href{ https://doi.org/10.1038/s41467-017-02775-8     }{ Nat. Commun.         \textbf{9   }, 363 (2018).    }
           
           
           
            \bibitem{50}
           Y. Wang, J. Li, S. Zhang, K. Su, Y. Zhou, K. Liao, S. Du, H. Yan, S.-L. Zhu, Efficient quantum memory for single-photon polarization qubits. 
           \href{ https://doi.org/10.1038/s41566-019-0368-8     }{ Nat. Photon.          \textbf{13   }, 346-351 (2019).    }
           
           
           \bibitem{51}
           M. Cao, F. Hoffet, S. Qiu, A. S. Sheremet, J. Laurat, Efficient reversible entanglement transfer between light and quantum memories.
           \href{ https://doi.org/10.1364/optica.400695      }{ Optica   \textbf{ 7  }, 1440 (2020).     }
           
           
           \bibitem{52}
            Z. Yan, L. Wu, X. Jia, Y. Liu, R. Deng, S. Li, H. Wang, C. Xie, K. Peng, Establishing and storing of deterministic quantum entanglement among three distant atomic ensembles.  
           \href{ https://doi.org/10.1038/s41467-017-00809-9     }{ Nat. Commun.         \textbf{8   }, 718 (2017).    }
           
           
           \bibitem{53}
            E. Saglamyurek, N. Sinclair, J. Jin, J. A. Slater, D. Oblak, F. Bussieres, M. George, R. Ricken, W. Sohler, W. Tittel, Broadband waveguide quantum memory for entangled photons. 
           \href{  https://doi.org/10.1038/nature09719     }{ Nature        \textbf{469   }, 512-515 (2011).    }
           
           
           \bibitem{54}
            C. Clausen, I. Usmani, F. Bussieres, N. Sangouard, M. Afzelius, H. de Riedmatten, N. Gisin, Quantum storage of photonic entanglement in a crystal.   
           \href{https://doi.org/10.1038/nature09662   }{ Nature \textbf{469   }, 508-511 (2011).    }
           
           
           \bibitem{55}
            E. Saglamyurek, J. Jin, V. B. Verma, M. D. Shaw, F. Marsili, S. W. Nam, D. Oblak, W. Tittel, Quantum storage of entangled telecom-wavelength photons in an erbium-doped optical fibre. 
           \href{https://doi.org/10.1038/nphoton.2014.311      }{ Nat. Photon.         \textbf{  9  }, 83-87 (2015).     }
           
           
           \bibitem{56}
            K. R. Ferguson, S. E. Beavan, J. J. Longdell, M. J. Sellars, Generation of Light with Multimode Time-Delayed Entanglement Using Storage in a Solid-State Spin-Wave Quantum Memory.  
           \href{ https://doi.org/10.1103/PhysRevLett.117.020501  }{  Phys. Rev. Lett.  \textbf{ 117 }, 020501 (2016). }
           
           
            \bibitem{57}
           Y. W. Cho, G. T. Campbell, J. L. Everett, J. Bernu, D. B. Higginbottom, M. T. Cao, J. Geng, N. P. Robins, P. K. Lam, B. C. Buchler, Highly efficient optical quantum memory with long coherence time in cold atoms.  
           \href{  https://doi.org/10.1364/optica.3.000100     }{   Optica      \textbf{ 3  }, 100-107 (2016).    }
           
           
           
            \bibitem{58}
            M. Sabooni, Q. Li, S. Kroll, L. Rippe, Efficient quantum memory using a weakly absorbing sample. 
           \href{  https://doi.org/10.1103/PhysRevLett.110.133604 }{ Phys. Rev. Lett.  \textbf{110}, 133604 (2013).  }
           
           
           
            \bibitem{59}
           N. Sinclair, E. Saglamyurek, H. Mallahzadeh, J. A. Slater, M. George, R. Ricken, M. P. Hedges, D. Oblak, C. Simon, W. Sohler, W. Tittel, Spectral multiplexing for scalable quantum photonics using an atomic frequency comb quantum memory and feed-forward control. 
           \href{ https://doi.org/10.1103/PhysRevLett.113.053603      }{Phys. Rev. Lett.       \textbf{ 113  }, 053603 (2014).       }
           
           
           
           
            \bibitem{60}
           M. Hosseini, G. Campbell, B. M. Sparkes, P. K. Lam, B. C. Buchler, Unconditional room-temperature quantum memory.
           \href{https://doi.org/10.1038/nphys2021      }{ Nat. Phys.          \textbf{  7  }, 794-798 (2011).   }
           
           
           
            \bibitem{61}
           K. F. Reim, J. Nunn, V. O. Lorenz, B. J. Sussman, K. C. Lee, N. K. Langford, D. Jaksch, I. A. Walmsley, Towards high-speed optical quantum memories.  
           \href{ https://doi.org/10.1038/nphoton.2010.30     }{Nat. Photon.  \textbf{ 4  }, 218 (2010).     }
           
           
           
            \bibitem{62}
           J. Guo, X. Feng, P. Yang, Z. Yu, L. Q. Chen, C. H. Yuan, W. Zhang, High-performance Raman quantum memory with optimal control in room temperature atoms. 
           \href{ https://doi.org/10.1038/s41467-018-08118-5     }{  Nat. Commun.        \textbf{10    }, 148 (2019).   }
           
           
           \bibitem{63}
           D.-S. Ding, W. Zhang, Z.-Y. Zhou, S. Shi, B.-S. Shi, G.-C. Guo, Raman quantum memory of photonic polarized entanglement. 
           \href{ https://doi.org/10.1038/nphoton.2015.43     }{  Nat. Photon.         \textbf{9 }, 332-338 (2015).   }
           
           
           \bibitem{64}
           K. T. Kaczmarek, P. M. Ledingham, B. Brecht, S. E. Thomas, G. S. Thekkadath, O. Lazo-Arjona, J. H. D. Munns, E. Poem, A. Feizpour, D. J. Saunders, J. Nunn, I. A. Walmsley, High-speed noise-free optical quantum memory.  
           \href{ https://doi.org/10.1103/PhysRevA.97.042316     }{Phys. Rev. A          \textbf{ 97  }, 042316 (2018).     }
           
           
           
           \bibitem{65}
           Ran Finkelstein, Eilon Poem, Ohad Michel, Ohr Lahad, O. Firstenberg, Fast, noise-free memory for photon synchronization at room temperature.  
           \href{https://doi.org/10.1126/sciadv.aap8598      }{ Sci. Adv. \textbf{4          }, eaap8598 (2018).      }
           
           
           \bibitem{66}
           C. Simon, H. de Riedmatten, M. Afzelius, N. Sangouard, H. Zbinden, N. Gisin, Quantum repeaters with photon pair sources and multimode memories. 
           \href{https://doi.org/10.1103/PhysRevLett.98.190503      }{ Phys. Rev. Lett.          \textbf{98  }, 190503 (2007).      }
           
           
           
           \bibitem{67}
           D. G. England, P. S. Michelberger, T. F. M. Champion, K. F. Reim, K. C. Lee, M. R. Sprague, X. M. Jin, N. K. Langford, W. S. Kolthammer, J. Nunn, I. A. Walmsley, High-fidelity polarization storage in a gigahertz bandwidth quantum memory.  
           \href{ https://doi.org/10.1088/0953-4075/45/12/124008     }{Journal of Physics B: Atomic, Molecular and Optical Physics        \textbf{45   }, 124008 (2012).       }
           
           
           
           \bibitem{68}
           Chin-Wen Chou, Julien Laurat, Hui Deng, Kyung Soo Choi, Hugues de Riedmatten, Daniel Felinto, H. J. Kimble, Functional quantum nodes for entanglement distribution over scalable quantum networks.
           \href{  https://doi.org/10.1126/science.1140300     }{  Science        \textbf{316    }, 1316-1319 (2007).     }
           
           
           
           \bibitem{69}
            K. C. Cox, D. H. Meyer, Z. A. Castillo, F. K. Fatemi, P. D. Kunz, Spin-Wave Multiplexed Atom-Cavity Electrodynamics. 
           \href{ https://doi.org/10.1103/PhysRevLett.123.263601     }{ Phys. Rev. Lett.        \textbf{123   }, 263601 (2019).       }
           
           
           
           \bibitem{70}
            L. Heller, P. Farrera, G. Heinze, H. de Riedmatten, Cold-Atom Temporally Multiplexed Quantum Memory with Cavity-Enhanced Noise Suppression.  
           \href{ https://doi.org/10.1103/PhysRevLett.124.210504      }{Phys. Rev. Lett.         \textbf{124   }, 210504 (2020).}
           
           
           
           
           \bibitem{71}
           See Supplemental Material.
           \href{      }{         \textbf{   }     }
           
           
           \bibitem{72}
           H. de Riedmatten, J. Laurat, C. W. Chou, E. W. Schomburg, D. Felinto, H. J. Kimble, Direct measurement of decoherence for entanglement between a photon and stored atomic excitation.  
           \href{https://doi.org/10.1103/PhysRevLett.97.113603      }{ Phys. Rev. Lett.        \textbf{ 97   }, 113603 (2006).     }     
           
           \bibitem{73}
            S. Chen, Y. A. Chen, B. Zhao, Z. S. Yuan, J. Schmiedmayer, J. W. Pan, Demonstration of a stable atom-photon entanglement source for quantum repeaters.
           \href{ https://doi.org/10.1103/PhysRevLett.99.180505      }{ Phys. Rev. Lett.       \textbf{99    }, 180505 (2007).     }
           
           \bibitem{74}
            I. Marcikic, H. de Riedmatten, W. Tittel, H. Zbinden, M. Legre, N. Gisin, Distribution of time-bin entangled qubits over 50 km of optical fiber. 
           \href{ https://doi.org/10.1103/PhysRevLett.93.180502  }{Phys. Rev. Lett.    \textbf{93   }, 180502 (2004).  }
           
           \bibitem{75}
           H. R. B. Markus Aspelmeyer, Tsewang Gyatso, Thomas Jennewein, Rainer Kaltenbaek, Michael Lindenthal, Gabriel Molina-Terriza, Andreas Poppe, Kevin Resch, Michael Taraba, Rupert Ursin, Philip Walther, Anton Zeilinger, Long-Distance Free-Space Distribution of Quantum Entanglement. 
           \href{ https://doi.org/10.1126/SCIENCE.1085593      }{ Science        \textbf{ 301   }, 621 (2003).     }
           
           \bibitem{76}
           B. Albrecht, P. Farrera, X. Fernandez-Gonzalvo, M. Cristiani, H. de Riedmatten, A waveguide frequency converter connecting rubidium-based quantum memories to the telecom C-band.  
           \href{https://doi.org/10.1038/ncomms4376      }{Nat. Commun.         \textbf{ 5   }, 3376 (2014).    }
           
      

      \end{thebibliography}

\begin{thebibliography}{9} 
      	\bibitem{1}
      	S.-Z. Wang, M.-J. Wang, Y.-F. Wen, Z.-X. Xu, T.-F. Ma, S.-J. Li, H. Wang, Long-lived and multiplexed atom-photon entanglement interface with feed-forward-controlled readouts. 
      	\href{https://doi.org/10.1038/s42005-021-00670-9      }{ Communications Physics         \textbf{ 4  }, 168 (2021).    }
      	
      	 \bibitem{2}
      	X.-H. Bao, A. Reingruber, P. Dietrich, J. Rui, A. Dück, T. Strassel, L. Li, N.-L. Liu, B. Zhao, J.-W. Pan, Efficient and long-lived quantum memory with cold atoms inside a ring cavity. 
      	\href{https://doi.org/10.1038/nphys2324  }{  Nat. Phys. \textbf{8 },517-521 (2012). }
      	
      	\bibitem{3}
      	S. Chen, Y. A. Chen, T. Strassel, Z. S. Yuan, B. Zhao, J. Schmiedmayer, J. W. Pan, Deterministic and storable single-photon source based on a quantum memory. 
        \href{https://doi.org/10.1103/PhysRevLett.97.173004  }{Phys. Rev. Lett.    \textbf{97}, 173004 (2006)}
        
        \bibitem{4}
        S.-J. Yang, X.-J. Wang, X.-H. Bao, J.-W. Pan, An efficient quantum light–matter interface with sub-second lifetime.    
        \href{https://doi.org/10.1038/nphoton.2016.51      }{Nat. Photon.  \textbf{ 10  },381-384 (2016). }
         \bibitem{5}
        S. J. Yang, X. J. Wang, J. Li, J. Rui, X. H. Bao, J. W. Pan, Highly retrievable spin-wave-photon entanglement source.  
        \href{https://doi.org/10.1103/PhysRevLett.114.210501      }{ Phys. Rev. Lett.       \textbf{114     }, 210501 (2015).}
         \bibitem{6}
        B. Zhao, Y.-A. Chen, X.-H. Bao, T. Strassel, C.-S. Chuu, X.-M. Jin, J. Schmiedmayer, Z.-S. Yuan, S. Chen, J.-W. Pan, A millisecond quantum memory for scalable quantum networks.  
        \href{https://doi.org/10.1038/nphys1153      }{ Nat. Phys. \textbf{5   }, 95-99 (2008).    }
        
         
         \bibitem{7}
         E. L. Adriana, J. M. Aaron, W. N. Sae, Counting near-infrared single-photons with 95\% efficiency.  
         \href{https://doi.org/10.1364/OE.16.003032  }{Opt. Express\    \textbf{16},3032 (2008). }
         
         \bibitem{8}
         C. Simon, H. de Riedmatten, M. Afzelius, N. Sangouard, H. Zbinden, N. Gisin, Quantum repeaters with photon pair sources and multimode memories. 
         \href{https://doi.org/10.1103/PhysRevLett.98.190503      }{ Phys. Rev. Lett.          \textbf{98  }, 190503 (2007).      }
         
         \bibitem{9}
         O. A. Collins, S. D. Jenkins, A. Kuzmich, and T. A. Kennedy,Multiplexed memory-insensitive quantum repeaters.  
         \href{https://doi.org/10.1103/PhysRevLett.98.060502      }{ Phys. Rev. Lett.          \textbf{98  }, 060502 (2007).}
         \bibitem{10}
            D. Lago-Rivera, S. Grandi, J. V. Rakonjac, A. Seri, H. de Riedmatten, Telecom-heralded entanglement between multimode solid-state quantum memories.   
         \href{https://doi.org/10.1038/s41586-021-03481-8      }{Nature           \textbf{594  }, 37-40 (2021).}
          \bibitem{11}
         X. Liu, J. Hu, Z. F. Li, X. Li, P. Y. Li, P. J. Liang, Z. Q. Zhou, C. F. Li, G. C. Guo, Heralded entanglement distribution between two absorptive quantum memories.  
         \href{https://doi.org/10.1038/s41586-021-03505-3      }{ Nature           \textbf{594  }, 41-45 (2021).}
         \bibitem{12}
         B. Zhao, Z. B. Chen, Y. A. Chen, J. Schmiedmayer, J. W. Pan, Robust creation of entanglement between remote memory qubits.  
         \href{https://doi.org/10.1103/PhysRevLett.98.240502      }{Phys. Rev. Lett.           \textbf{ 98  }, 240502 (2007).}
         
         
         \bibitem{13}
         Y. Yu, F. Ma, X. Y. Luo, B. Jing, P. F. Sun, R. Z. Fang, C. W. Yang, H. Liu, M. Y. Zheng, X. P. Xie, W. J. Zhang, L. X. You, Z. Wang, T. Y. Chen, Q. Zhang, X. H. Bao, J. W. Pan, Entanglement of two quantum memories via fibres over dozens of kilometres.  
         \href{ https://doi.org/10.1038/s41586-020-1976-7      }{Nature         \textbf{578   }, 240-245 (2020).     }
          \bibitem{14}
         Y. O. Dudin, L. Li, A. Kuzmich, Light storage on the time scale of a minute. 
         \href{https://doi.org/10.1103/PhysRevA.87.031801      }{ Phys. Rev. A  (2013).           \textbf{ 87 }, 031801(R)}
         
         \bibitem{15}
         M. Lipka, M. Mazelanik, A. Leszczyński, W. Wasilewski, M. Parniak, Massively-multiplexed generation of Bell-type entanglement using a quantum memory.  
         \href{https://doi.org/10.1038/s42005-021-00551-1      }{Communications Physics            \textbf{4  }, 46 (2021).}
         
          \bibitem{16}
          Y. Wen, P. Zhou, Z. Xu, L. Yuan, H. Zhang, S. Wang, L. Tian, S. Li, H. Wang, Multiplexed spin-wave–photon entanglement source using temporal multimode memories and feedforward-controlled readout.   
         \href{https://doi.org/10.1103/PhysRevA.100.012342      }{Phys. Rev. A           \textbf{100  }, 012342 (2019).}
          \bibitem{17}
        N. Sinclair, E. Saglamyurek, H. Mallahzadeh, J. A. Slater, M. George, R. Ricken, M. P. Hedges, D. Oblak, C. Simon, W. Sohler, W. Tittel, Spectral multiplexing for scalable quantum photonics using an atomic frequency comb quantum memory and feed-forward control.  
         \href{https://doi.org/10.1103/PhysRevLett.113.053603      }{Phys. Rev. Lett.            \textbf{113  }, 053603 (2014).}
          \bibitem{18}
         T. S. Yang, Z. Q. Zhou, Y. L. Hua, X. Liu, Z. F. Li, P. Y. Li, Y. Ma, C. Liu, P. J. Liang, X. Li, Y. X. Xiao, J. Hu, C. F. Li, G. C. Guo, Multiplexed storage and real-time manipulation based on a multiple degree-of-freedom quantum memory.  
         \href{https://doi.org/10.1038/s41467-018-05669-5      }{ Nat. Commun.           \textbf{9  }, 3407 (2018).}
         
         
         
        
          
         
\end{thebibliography}

\end{document}